# Precipitate phase selection and grain boundary morphology in Cu-Ni-Si-Mn alloys: A machine-learning interatomic potential study

Aadil Fayaz Wani,[1] Il-Seok Jeong,[2] Haekwan Jeon,[3] Jaesun Kim,[3] SuDong Park,[1] Eun-Ae Choi,[2,4*] Seung Zeon Han,[2,4] Seungwu Han,[3,5] and Byungki Ryu[1,6*]

1- Energy Conversion Research Center, Electrical Materials Research Division, Korea Electrotechnology Research Institute (KERI), Changwon-si 51543, Republic of Korea
2- Extreme Materials Research Institute, Korea Institute of Materials Science (KIMS), Changwon-si 51508, Republic of Korea
3- Department of Materials Science and Engineering, Seoul National University, Seoul 08826, Republic of Korea
4- Advanced Materials Engineering, KIMS School, University of Science and Technology (UST), Changwon-si 51508, Republic of Korea
5- AI Center, Korea Institute for Advanced Study, Seoul 02455, Republic of Korea
6- Electrical-Functionality Materials Engineering, KERI School, University of Science and Technology (UST), Changwon-si 51543, Republic of Korea

*Correspondence: byungkiryu@keri.re.kr (BR) eunae.choi@kims.re.kr (EA)

**Abstract**

Alloys inevitably contain interphase boundaries, whose energetics govern nucleation processes and precipitate morphology. In Cu-Ni-Si alloys, Mn addition markedly changes grain boundary (GB) precipitation behavior. While GB precipitation of stable $Ni_2Si$ in Mn-free alloys is associated with degraded mechanical properties, Mn addition instead promotes film-shaped $Mn_6Ni_{16}Si_7$ (G-phase) precipitation, which is correlated to improved mechanical properties. However, the atomic origin of the contrasting GB phase selection and morphology remains unclear. Here we perform machine-learning interatomic potential (MLIP) calculations to investigate the effect of interphase-boundary atomic structure on GB precipitates in Cu-Ni-Si alloys with and without Mn. The MLIP calculations reliably reproduce DFT-level energetics for interfacial bonding and microstructural configurations, and further predict that Mn addition favors GB precipitation of $Mn_6Ni_{16}Si_7$ rather than $Ni_2Si$. Experimentally, Mn-free alloys are observed to exhibit irregularly shaped $Ni_2Si$ precipitates with open-boundary-like Cu/$Ni_2Si$ interfaces, whereas Mn-added alloys exhibit film-like G-phase at GBs. Large-scale atomistic interface calculations reveal that the coherent interface structure between Cu and $Ni_2Si$ favors the formation of plate-like strained $Ni_2Si$ precipitates in the matrix. Upon coarsening and stress release, an out-of-phase coherent-like interface can form at GBs, generating a local repulsive region that gives rise to surface-like open-boundary structures and explains the irregular morphology of GB stable $Ni_2Si$ precipitates. In contrast, Cu/$Mn_6Ni_{16}Si_7$ interfaces remain predominantly incoherent with moderate boundary energies and no pronounced repulsive regime, thereby stabilizing continuous interfacial contact and explaining film-shaped GB precipitation.

**Keywords**

Cu-Ni-Si-Mn alloys; Precipitate morphology; Phase stability; Grain boundary precipitation; Machine learning interatomic potentials

## 1. Introduction

Cu alloys are widely used for their high electrical conductivity [1–3]. Although pure Cu exhibits excellent conductivity, it is mechanically soft and has low strength. Precipitation hardening has attracted sustained interest because controlled formation of nanoscale precipitates in the Cu matrix increases mechanical strength without severely degrading electrical conductivity[4,5]. In particular, strained metastable $\delta$-$Ni_2Si$ precipitates are widely recognized as the primary strengthening phase in Cu-Ni-Si alloys. The precipitation of finely dispersed $Ni_2Si$ during aging after solid solution treatment significantly enhances the strength of Cu alloys by impeding dislocation motion[6,7]. Moreover, the coherent interfaces between matrix and precipitate can preserve relatively high conductivity without introducing excessive electron scattering by defects at boundaries[8,9].

A high volume fraction of $Ni_2Si$ particles in the matrix can increase alloy strength, making it useful for many applications [10–13]. One simple way to increase the amount of these strengthening particles is to add more Ni and Si elements. However, when the Ni and Si contents are too high, stable and non-strained particles may form and grow at grain boundaries (GBs) [14,15]. As a result, Cu-Ni-Si alloys with high Ni and Si contents are difficult to process by methods such as rolling and drawing. Therefore, these alloys are more suitable for use in the cast condition rather than as wrought materials. Alloying additional elements such as Co further refines precipitation behavior, promoting the nucleation of $(Ni,Co)_2Si$ platelets with reduced interfacial energy and improved thermal stability [16–19]. This ultimately results in superior mechanical performance and precipitation kinetics in Cu-Ni-Si alloys.

Recent density functional theory (DFT) calculations suggested that the addition of 3d transition metals may control precipitation stability in Cu by modulating precipitate/matrix interface energies[20]. While Ti, V, Cr, Mn, Fe, and Co were found to stabilize $Ni_2Si$ cohesive energy, only Mn atoms can decrease the interfacial energy. A recent study reported that Mn-added Cu-Ni-Si exhibit a richer precipitation landscape, where the interplay of solute interactions can stabilize more complex structures such as $Mn_6Ni_{16}Si_7$ grain boundary phase (G-phase) [21–25]. In Fe-based alloys, numerous experimental studies have reported that G-phase precipitates causes degraded mechanical responses: ferrite embrittlement and loss of hardening [26–28]. In contrast, Han et al. reported an enhanced mechanical strength in Mn-added Cu-Ni-Si alloys, associated with the heterogeneous nucleation of film-shaped G-phase precipitates at GBs [24]. However, the atomic origin of this contrasting effect of G-phase precipitates remains unclear. In particular, why G-phase is detrimental in Fe-based alloys but beneficial in Cu-based alloys, and how the interphase boundary between the G-phase and Cu grains is structured, have not yet been quantitatively clarified.

Existing studies predominantly rely on experimental observations of precipitate morphology, distribution, and hardness evolution, or often supported by DFT calculations of bulk formation energies only. While such approaches successfully identify thermodynamically favorable phases, they are inherently limited in the ability of DFT calculations to describe precipitate stability within and outside the matrix. Precipitate selection is governed not only by chemical driving forces of bulk formation energy but also by interphase interface energetics, all of which depend strongly on precipitate size, shape, and coherency [29]. Consequently, experimentally observed transitions, such as the replacement of $\delta$-$Ni_2Si$ by G-phase precipitates upon Mn addition, are frequently rationalized qualitatively, without a unified energetic framework capable of predicting critical sizes or stability crossovers.

A major challenge in addressing this gap is the length-scale limitation of first-principles methods. DFT calculations are restricted to small supercells less than 1,000 atoms and therefore cannot directly capture realistic precipitate dimensions larger than 1-10 nm or systematically explore size-dependent energetics. As a result, previous computational studies have largely focused on bulk or surface energies, leaving the combined effects of bulk, interface, surface, and strain energies unresolved [30–32]. This limitation has hindered the development of predictive models for precipitate selection in multicomponent Cu alloys.

Recent advances in machine learning interatomic potentials (MLIPs) provide a promising route to overcome the challenges mentioned above [33–40]. MLIPs make it possible to simulations at nanometer length scales and evaluate complex energetic contributions for realistic precipitate geometries and sizes while retaining accuracy close to DFT[41,42]. However, despite rapid progress in MLIP development, their reliability for describing subtle energetic competition between different intermetallic precipitates in multicomponent alloys remains largely unexplored. In particular, it is not yet established whether MLIPs can faithfully reproduce DFT-level energetics trends in phase stability, interfacial energetics, and strain energy quantities that are critical for precipitation stability and kinetics.

In this work, we address these challenges by developing a quantitative, size-dependent energetic framework to rationalize precipitate phase selection in Cu-Ni-Si-Mn alloys. Using a combination of DFT and state-of-the-art MLIP calculations, we systematically decompose the total energy of $\delta$-$Ni_2Si$ and G-phase precipitates in Cu alloys into bulk, interface, and strain energy contributions. For heterogeneous nucleation at grain boundaries, we also consider the surface energies as a limiting value. This approach enables direct comparison of competing precipitate phases as a function of size and matrix constraint, allowing identification of critical stability crossovers between precipitates and nucleation mode so that one can access DFT-level description of precipitates and related system. By showing that MLIP calculations consistently reproduce DFT trends across all relevant energetic terms, this study establishes the reliability of MLIPs for predictive modeling of complex energetics of precipitates including size-dependent particle stability in alloys.

The present results provide mechanistic insight into the origin of precipitation phase selection and precipitate morphology in Cu-Ni-Si alloys with and without Mn addition by demonstrating that interfacial stability dictates the competitive growth of specific precipitate phases rather than bulk energetics alone. Our quantitative analysis reveals that the interface and strain energies act as the primary filters for precipitate survival within the Cu matrix. Furthermore, this work offers a transferable computational framework that leverages DFT-validated MLIPs to bridge the gap between atomic-scale interfacial structures and the microstructural-scale stability of multicomponent metallic systems.

The MLIP calculations also enable direct atomistic simulations of interfaces at length scales beyond a few nm. We predict strain-released Cu/$Ni_2Si$ interface structures using MLIP calculations. Strained metastable $Ni_2Si$ is favored by a coherent Cu/$Ni_2Si$ interface, which supports precipitate formation in the Cu matrix. However, as the precipitate coarsens and elastic strain is released, this coherent configuration can evolve into an out-of-phase coherent-like interface that creates a local repulsive region. Such a repulsive regime promotes the formation of a gapped, open-boundary-like interfacial structure, thereby explaining the irregular morphology of $Ni_2Si$ precipitates at grain boundaries, in good agreement with the experimental observations. In contrast, the Cu/$Mn_6Ni_{16}Si_7$ interface remains predominantly incoherent, with moderately low interfacial energies and no pronounced repulsive regime, which accounts for the continuous contact and film-shaped morphology of the G-phase.

## 2. Methods

### 2.1 DFT calculation.

First-principles calculations were carried out within the DFT framework [43,44]. The exchange-correlation effects were described using the generalized gradient approximation (GGA), employing the Perdew-Burke-Ernzerhof (PBE) functional [45], as implemented in the Vienna ab initio Simulation Package (VASP) [46]. The interaction between valence and core electrons was treated using the projector-augmented wave method (PAW), ensuring an accurate and efficient description of the electronic structure [47]. The plane-wave basis set was defined using a kinetic energy cutoff of 560 eV, 460 eV, and 500 eV for cubic Cu, orthorhombic $Ni_2Si$, and cubic $Mn_6Ni_{16}Si_7$ compounds, respectively, to ensure convergence. To maintain consistency across the interface models, a uniform cutoff of 560 eV was used. For pristine FCC-Ni, FCC-Si, $\alpha$-Mn, and $\beta$-Mn similar energy cutoff of 500 eV is used to avoid any inconsistence in energy convergence. For all interface calculations, a uniform cutoff of 560 eV was adopted. The Brillouin zone was sampled using Monkhorst-Pack k-point grids of 4×4×4 for the bulk structures and 2×2×2 for interfaces. Electronic occupations were treated using Gaussian smearing is used with energy width of 0.01 eV to ensure numerical stability. Structural relaxations were performed until the systems reached a fully relaxed state, defined by force convergence criteria of $10^{-3}$ eV/Å.

### 2.2 MLIP calculation setting.

We performed MLIP calculations to evaluate structure energetics of bulk, interface, and surface structures. In this study, we employ pretrained MLIPs that differ in both their model architectures and pretraining datasets. Specifically, we compare M3GNet[48] and SevenNet[49] (hereafter referred to as 7net), which belong to the classes of E(3)-invariant and E(3)-equivariant graph neural networks, respectively. M3GNet relies on E(3)-invariant geometric descriptors, such as bond lengths and angles, whereas 7net additionally incorporates higher-order E(3)-equivariant tensor features in its latent representations. In both approaches, the predicted potential energy is invariant under relevant symmetry operations, while atomic forces and cell stress are obtained from derivatives of the energy, providing consistency among energy, force, and stress predictions.

For the M3GNet-based MLIP, we use a model pretrained on the MPF.2021.2.8 database, which consists of subsampled geometry optimization trajectories of inorganic crystalline materials. In the case of 7net, we consider three pretrained variants: **7net-0** [49], **7net-Ompa**, and **7net-Omni** [50]. The 7net-0 model is pretrained on the MPtrj database [51], which is derived from similar sources as MPF.2021.2.8 but is roughly an order of magnitude larger by updating materials and sampling methods. The 7net-ompa and 7net-Omni models further extend the training data to include more diverse inorganic compositions and configurations, while 7net-Omni additionally incorporates low-dimensional systems such as surfaces and molecules. Notably, both 7net-Ompa and 7net-Omni adopt a multi-fidelity framework [52] that enables simultaneous training on datasets generated with varying computational settings. In this work, we examine several inference modes of the multi-fidelity 7net models, corresponding to different DFT levels used during pretraining. In particular, we focus on the mpa and omat24 channels (as defined in Ref. [52]), which are tailored for bulk material simulations.

Bulk, interface, and surface configurations as well as isolated single-solute state were fully relaxed within each MLIP framework with force convergence criteria ($f_{max}$) of 0.001 eV/Å, and then their energies were computed and compared against DFT.

### 2.3 Atomic chemical potentials and formation energies.

**Phase formation energy.** Formation energy per atom ($E_{\mathrm{f}}$) is calculated relative to elemental reference states under identical computational settings, enabling a consistent comparison of the thermodynamic driving forces for phase formation. The $E_{\mathrm{f}}$ is calculated using the relation:

$$E_{\mathrm{f}} = \frac{1}{x+y+z}\left\{E_{\mathrm{tot}}\left(\mathrm{A}_x\mathrm{B}_y\mathrm{C}_z\right) - \left[x \cdot E_{\mathrm{bulk}}(\mathrm{A}_x) + y \cdot E_{\mathrm{bulk}}\left(\mathrm{B}_y\right) + z \cdot E_{\mathrm{bulk}}(\mathrm{C}_z)\right]\right\} \quad (1)$$

where $E_{\mathrm{tot}}\left(\mathrm{A}_x\mathrm{B}_y\mathrm{C}_z\right)$ is total bulk energy of $\mathrm{A}_x\mathrm{B}_y\mathrm{C}_z$ compound containing $x$, $y$, $z$ atoms and $E_{\mathrm{bulk}}$ is the total energy per atom of the constituent A, B, C elements in their bulk stable form. This energy corresponds to the energy of formation when the pure elemental components are combined to form the bulk phase.

**Solute chemical potentials in the Cu matrix.** The chemical potential of solute X (X = Mn, Si, and Ni) is calculated with reference to matrix chemical potential (Cu) as:

$$\mu_{\mathrm{solute}}(\mathrm{X}) = E[\mathrm{Cu}_{n-1}\mathrm{X}] - E[\mathrm{Cu}_n] + \mu(\mathrm{Cu}) \quad (2)$$

and

$$\mu(\mathrm{Cu}) = \frac{1}{n}E[\mathrm{Cu}_n]. \quad (3)$$

where $E[\mathrm{Cu}_n\mathrm{X}]$ is the total bulk energy of fcc Cu containing solute atom A and $n$ are the total number of Cu atoms. The chemical potential of Cu is represented by $\mu(\mathrm{Cu})$.

To compare the relative formation energy of solutes in Cu alloy, solid-solution configurations containing Ni, Mn, and Si solutes in FCC Cu were constructed using Cu (3 × 3 × 3) cubic supercells to minimize solute-solute interactions. The bulk energy of the precipitate phase within the Cu matrix is defined relative to the chemical potentials of its constituent elements in solute states, as follows:

$$E_{\mathrm{bulk}}[\mathrm{precipitate}] = \left(E_{\mathrm{tot}} - \sum_i n_i\mu_i\right) \quad (4)$$

where $E_{\text{tot}}$ is the total energy of the Cu-matrix supercell containing the solute/precipitate, $n_i$ represents the number of atoms of element $i$, and $\mu_i$ is the chemical potential (energy per atom) of element $i$ in its respective reference state. The total number of atoms in the supercell is denoted by $N$, ensuring the formation energy is normalized per atom.

Different solute atoms are used to assess the possibility of forming a stable solid solution. Mono- and dimer-Mn are single and double Mn solute atoms in Cu, while $\alpha$ and $\beta$-Mn are $Mn_{58}$ and $Mn_{20}$ structures used for calculation of Mn solute reference energy. These are the primary solid-phases of Manganese with $\alpha$-Mn stable at low temperature while β-Mn at high temperature. Additionally, FCC $Ni_2Si$ and $\delta$-$Ni_2Si$ correspond to the ideal anti-fluorite and orthorhombic $Ni_2Si$ structures, respectively.

**Bulk energy contribution to precipitate formation.** In solid-solution Cu alloys, solute atoms may cluster and form a precipitate. However, precipitate formation is favorable only when the bulk formation driving force exceeds the accompanying surface and interfacial energy penalties; otherwise, the solutes remain dissolved in the Cu matrix. To capture this competition, all total energies are defined relative to the bulk solid-solution reference state. Note that this shift does not affect the values of the heat of formation or the solute chemical potentials.

### 2.4 Atomic models and their energetics for strained phase, interface, and surface structures

To perform reliable calculations, the interface plane of two constituent materials must possess equal planer dimensions. However, due to the mismatch between planer dimensions of Cu and precipitate phases, the interface is not possible without structural adjustments. To achieve lattice compatibility, an in-plane strain is imposed on the precipitate to match the Cu matrix. The energetic cost associated with this deformation is defined as the strain energy ($E_{\text{str}}$), which corresponds to the energy difference between the relaxed equilibrium structure and the strained configuration, and is calculated as:

$$E_{\text{str}} = E_{(\text{strained A})} - E_{(\text{equilibrium A})} \tag{5}$$

where $E_{(\text{equilibrium A})}$ and $E_{(\text{strained A})}$ are the bulk energy of relaxed and deformed structures, respectively[53]. For $Ni_2Si$, the (100) plane of $Ni_2Si$ was coherently matched with the $(1\bar{1}0)$ plane of Cu by applying a controlled lattice distortion to $Ni_2Si$. Similarly, for the G-phase, the (100) plane was aligned with the Cu (100) plane by applying a small strain to the G-phase along the same plane.

Once the Cu and precipitate structures are designed, commensurate interface models are constructed separated by sufficiently thick bulk regions (a few Å) to suppress interface-interface interactions. Atomic positions were fully relaxed while keeping the in-plane lattice parameters fixed while the interface supercell lattice parameter normal to the interface is optimized. The interface energy ($E_{\mathrm{IF}}$) is calculated as:

$$E_{\mathrm{IF}} = \frac{E_{(\text{strained A on Cu})} - (E_{(\text{strained A})} + E_{(\text{bulk Cu})})}{S} \tag{6}$$

where $E_{(\text{strained A})}$ , $E_{(\text{bulk Cu})}$ , and $E_{(\text{strained A on Cu})}$ are the bulk total energies of the strained precipitate (A = $Ni_2Si$ or $Mn_6Ni_{16}Si_7$), bulk Cu and interface containing Cu and strained A [54,55]. Additionally, $S$ is an interface area which is taken two times the area of single interface to account for the two interfaces in the coherent interface model.

Surface energies ($E_{\mathrm{SF}}$) are evaluated using slab models constructed from fully relaxed bulk structures. A sufficiently thick vacuum layer of about 15 Å was inserted normal to the surface to avoid interactions between periodic images. During relaxation, atomic positions were optimized while maintaining the in-plane lattice parameters fixed to their bulk values. The $E_{\mathrm{SF}}$ was calculated as:

$$E_{\mathrm{SF}} = \frac{E_{(\text{slab A})} - E_{(\text{bulk A})}}{S}. \tag{7}$$

Here $E_{(\text{bulk A})}$ and $E_{(\text{slab A})}$ are the total energy of bulk and slab structures, respectively and $S$ is the total surface area of two surfaces in the slab structure [56,57].

## 2.5 Precipitate energy modelling

The total energy of alloy system with a single precipitate can be expressed as the combined contribution of bulk, strain, interface and surface energy terms. These contributions are inherently dependent on precipitate size and collectively determine the stability of precipitate inside and outside the Cu matrix.

The total system energy ($E_{\mathrm{sys}}(d)$) is written and defined as:

$$E_{\mathrm{sys}}(d) = (E_{\mathrm{bulk}} + E_{\mathrm{str}}) \cdot \alpha \cdot d^3 + E_{\mathrm{IF}} \cdot \beta \cdot d^2. \tag{8}$$

The constants $\alpha$ and $\beta$ are geometric factors associated with the volume and surface area of the precipitate, respectively and $d$ is a size variable (length or radius) for precipitate. The energy terms are computed with respect to solute states and are already defined. To analyze the size-dependent precipitate stability, the energy density is evaluated as [58,59]:

$$E_{\mathrm{den}}(d) = \frac{E_{\mathrm{sys}}}{\alpha \cdot d^3} = (E_{\mathrm{bulk}} + E_{\mathrm{str}}) + E_{\mathrm{IF\ or\ SF}} \cdot \left(\frac{\beta}{\alpha}\right) \cdot \frac{1}{d} \tag{9}$$

Since the bulk contribution is energetically favorable, the energy density can also be written as:

$$E_{\text{den}}(d) = -|E_{\text{bulk}}| + E_{\text{str}} + E_{\text{IF or SF}} \cdot \left(\frac{\beta}{\alpha}\right) \cdot \frac{1}{d} \tag{10}$$

The critical precipitate size $d_{\text{c}}$ is obtained from the stability condition:

$$E_{\text{den}}(d_{\text{c}}) = 0 \tag{11}$$

which defines the minimum size required for a precipitate to become energetically stable within the Cu matrix.

## 3. Results

### 3.1 Atomic models and reference energetics: DFT calculations

From the experimental data, it is found that both Cu and G-phase ($Mn_6Ni_{16}Si_7$) are stable in cubic structure while $Ni_2Si$ stabilize in orthorhombic structure[60]. To construct commensurate interfaces between the Cu matrix and $Ni_2Si$ precipitate, both the structures were transformed into an orthorhombic representation to minimize lattice mismatch. The Cu matrix was reoriented into an orthorhombic cell where the $[1\bar{1}0]$ and $[110]$ directions align with the x and y axes, respectively. This transformation results in a Cu orthorhombic structure with base dimensions of $\sqrt{2}a$, where $a$ is lattice parameter of fcc Cu. To ensure a commensurate match, the reoriented Cu was subsequently expanded to $3\sqrt{2}a \times \sqrt{2}a \times a$ supercell and $Ni_2Si$ lattice was extended by two units along the x-axis creating a $2 \times 1 \times 1$ supercell. Similarly, to minimize the computational cost while maintaining structural accuracy, a commensurate interface between the Cu matrix and the G-phase were constructed by transforming both lattices into a reduced orthorhombic structure. This yields a squared x-y basal plane with dimensions of approximately 7.67 Å. Supercell of 1×1×3 is used for Cu to ensure minimum volume mismatch as shown in **Figure 1**. These orientation relationships minimize the in-plane lattice misfit and enable the formation of commensurate interfaces. All interface structures were subsequently fully relaxed using DFT to eliminate residual forces and stresses, allowing the systems to reach mechanically stable configurations under the imposed coherency constraint. The fully optimized structural parameters used for all subsequent simulations are listed in Tables S1 and S2.

Once the structures are designed, the heat of formation of $Ni_2Si$ and G-phase is calculated using the DFT framework to access their intrinsic thermodynamic stability. Both the phases are found stable in their optimized structural form with respect to their constituent elemental reference states. The calculated values of $E_{\text{f}}$ are -0.482 eV/atom and -0.616 eV/atom for $Ni_2Si$ and G-phase, respectively, as given in **Table 1**. The lower $E_f$ of G-phase indicates a stronger thermodynamic driving force for its formation under equilibrium conditions and suggests that G-phase is the energetically preferred phase against $Ni_2Si$ with reference to their elemental states.

While bulk formation energies provide insight into the intrinsic thermodynamic preference of the individual phases, the relative stability of $Ni_2Si$ and the G-phase within the Cu matrix must be evaluated with respect to Mn, Ni and Si solute states under realistic alloying conditions. We consider the chemical transformation from $Ni_2Si$ to G-phase, where excess Mn, Ni, or Si atom is balanced by the corresponding solute states or elemental phases, as follows:

$$8\ \mathrm{Ni_2Si} + 6\ \mathrm{Mn} \rightarrow \mathrm{Mn_6Ni_{16}Si_7} + \mathrm{Si} \tag{12}$$

Thus, there is a competition between $Ni_2Si$ and G-phase formation inside Cu matrix. To elucidate the thermodynamic driving forces governing the precipitation sequence in Cu-Ni-Si-Mn alloys, the relative formation energetics of $Ni_2Si$ and G-phase related configurations were systematically evaluated with respect to Cu solid solution reference. **Figure 2** presents the DFT relative energies associated with the transformation of $Ni_2Si$ and Mn into G-phase under various structural and chemical configurations. Since $\delta$-$Ni_2Si$ precipitate is well studied through theory and experiments and is found to form readily inside the Cu matrix during the early stages of aging, it is taken as a reference state to check the relative stability of other possible precipitates [20,24,61]. Notably, all precipitates are higher in energy except G-phase + Si configuration, which displays comparably lower energy than $Ni_2Si$-based states. This demonstrates that G-phase formation becomes thermodynamically competitive once sufficient solute redistribution occurs. This energetic hierarchy suggests that $Ni_2Si$ precipitation is favored at early stages, whereas G-phase formation is stabilized at later stages as diffusion and chemical ordering progress. It should be noted that these energies correspond only the bulk contribution without strain energy contribution, whereas the overall stability of precipitates strongly depends on its size-induced strain and boundary energy effect during nucleation and growth.

### 3.2 DFT benchmarked MLIPs: bulk, strain, interface, and surface energies

Reliable evaluation of precipitation energetics in multicomponent Cu-Ni-Si-Mn alloys critically depends on an accurate description of the equilibrium crystal structures of both the matrix and intermetallic phases. Since deviations in lattice parameters directly affect various energy terms and final results, we first benchmark the performance of the M3GNet and 7net MLIPs against DFT based structural reference data.

The DFT-optimized structures were re-optimized using various MLIP models to determine their corresponding equilibrium configurations. Since we have solely used different pretrained models which differ in their structural diversity of the training data, the resulting lattice parameters of Cu, $Ni_2Si$, and G-phase differ across each MLIP model as listed in **Table S1** and **S2**. For all phases, the MLIPs reproduce equilibrium lattice parameters and unit-cell volumes with high fidelity, exhibiting deviations of less than 2% relative to the DFT data. This validated accuracy establishes a reliable basis for subsequent calculations of formation energies, interface energetics, and size-dependent stability of $Ni_2Si$ and G-phase precipitates, discussed in subsequent sections.

As shown in **Figure 2**, all pretrained MLIP models reproduce the DFT energetic ordering among competing $Ni_2Si$ and G-phase-related configurations. In scenarios characterized by Mn-poor conditions, often resulting from Mn element phase separation, the calculations indicate that the $\delta$-$Ni_2Si$ phase can achieve thermodynamic stability. Conversely, under Mn-rich and Ni-rich conditions, the G-phase emerges as the more stable configuration with the lowest energy formation energy. This close quantitative and qualitative agreement demonstrates that the MLIPs accurately reproduce the energetic hierarchy obtained from first-principles calculations, including the stabilization of the G-phase upon sufficient solute redistribution. The excellent agreement across different MLIP architectures confirms their robustness in capturing subtle energy differences among competing precipitates within the Cu matrix.

Minor deviations are observed for 7net-Omni-mpa and 7net-Omni-omat models in the G-phase + mono-Si and G-phase + solute-Si configurations. The deviations are likely attributed to the fact that this specific model was pretrained using the OMAT dataset, which may capture a different subspace of the potential energy surface. However, the overall energetic trends and stability ordering remain fully consistent with DFT predictions. The slightly higher relative energies predicted by MLIPs can be attributed to a modest overestimation of solute formation energies, a known limitation of data-driven interatomic potentials. Nevertheless, the strong agreement with DFT indicates that MLIPs reliably capture the thermodynamic competition between $\delta$-$Ni_2Si$ and G-phase precipitates, enabling efficient and accurate large-scale simulations of precipitation behavior in Cu-Ni-Si-Mn alloys.

At early stages, the nucleation of $\delta$-$Ni_2Si$ and G-phase can start either inside the Cu matrix or at the grain boundary. The tendency of nucleation inside and outside of the matrix depends on different competent parameters. Inside the matrix, nucleation depends on bulk formation energy, elastic strain energy, and interface energy, while at the grain boundary, it depends on bulk formation energy and surface energy terms only. Hence, these parameters are calculated using DFT and MLIPs and are listed in **Table S3**. A close agreement is observed between DFT and MLIP results. The volume change, when $Ni_2Si$ or G-phase is distorted to match the Cu surface, is greater in $Ni_2Si$ than G-phase because the former makes a semi-coherent interface structure with Cu matrix, while the latter makes a coherent structure. This results in low elastic strain energy in G-phase than $Ni_2Si$. The surface energy of G-phase is higher than $Ni_2Si$, which is verified by all MLIPs. Since M3GNet and 7net are trained on MPtrj and Omat24 databases, we have chosen the same DFT settings as used by them, which leads to the close agreement of data between DFT and MLIPs.

It is noteworthy that MLIPs reproduce surface and interface energy trends in comparison to DFT values. The surface energy is defined as the energy difference between the relaxed bulk structure and the slab structure. Although surface configurations are less explicitly represented in the MLIP training dataset compared to bulk structures, the calculated surface energies exhibit less deviation compared to the strain energy, as shown in **Figure 3(a)**. This behavior arises primarily from error cancellation in the surface energy evaluation, which includes surface slab and bulk reference. Since the majority of atoms in the slab retain bulk-like coordination, and MLIPs accurately reproduce bulk energetics, systematic energy errors largely cancel out in the surface energy calculation. Furthermore, surface relaxation minimizes force-related inaccuracies, leading to similar relaxed configurations in both DFT and MLIP calculations. Consequently, despite limited explicit surface training, except Omni models, the MLIP provides reliable surface energy predictions. The error is slightly higher in case of M3GNet model due to the limited training dataset compared to other models.

Similarly, the MLIPs exhibit high accuracy in predicting interface energy, which includes the total energy of the supercell containing the interface, and reference bulk energies of Cu and $Ni_2Si$/G-phase. **Figure 3(b)** compares the values of interface energy obtained using DFT and other MLIP models. Since interface energy estimation includes only bulk reference energies, the high fidelity of MLIP predictions can be attributed to the extensive, consistent, and high-quality bulk configurations included in the training database, which enables accurate learning of bulk energetics. The DFT based minimum interface energy of Cu($1\bar{1}0$)/$Ni_2Si$(100) and Cu(100)/G-phase(100)interfaces are 0.48 J/m$^2$ and 0.98 J/m$^2$. The relative ease of epitaxial compatibility of atoms for $Ni_2Si$ results in a lower interface energy compared to the G-phase, which has a more disordered atomic configuration at the interface. Again, the error is relatively higher in the M3GNet model due to the low-fidelity dataset in the training model.

It should be noted that, although MLIPs are effective in estimating surface and interface energies, they are less effective in capturing the strain energy of a material. The strain energy is defined as the energy difference between the relaxed equilibrium structure and the deformed structure. Compared to other energy terms, the error in estimating strain energy is higher, as can be seen in **Figure 3 (c)**. This behavior is attributed to the fact that strain energy depends on the second derivatives of the potential energy surface (elastic constants), which are inherently more sensitive to small errors in the learned energy curvature. In contrast, surface and interface energies are governed primarily by local bonding environments, which are well represented in the training dataset. Consequently, small deviations in estimating elastic constants lead to higher error in the strain energy value.

### 3.3 Precipitate energetics in Cu alloys, E(d), and Size dependent stability of precipitates

**Figure 4** compares the size-dependent energy density and total energy of $Ni_2Si$ precipitates inside and outside the Cu matrix obtained from DFT with predictions from M3GNet and 7net-Ompa-mpa models. Data plots from DFT, M3GNet and one 7net model are used here for comparision while the data from other models are shown in **Figure S1** of supplementary materials. From DFT data, a nucleation barrier is observed at the grain boundary, while this barrier is negligible inside the matrix. A similar kind of nucleation barrier is captured by all the MLIP models with a small variation in the value of the energy hump. With increasing size, the energy density of both configurations converges toward a similar asymptotic value, corresponding to the bulk limit where surface and interface contributions become negligible. Both M3GNet and 7net-Ompa-mpa accurately reproduce this size-dependent crossover and the relative energetic ordering between inside and outside matrix configurations. This demonstrates a good transferability of MLIPs across different structural environments. Therefore, the magnitude and curvature of energy-size relationships are consistently captured, with only minor quantitative deviations. Importantly, the MLIPs accurately reproduce the absence of spurious energetic discontinuities or artificial nucleation barriers, demonstrating stable and transferable performance across different precipitate environments. The MLIPs closely reproduce the DFT trends across the entire particle-size range, including the relative energetic ordering between inside and outside matrix configurations and the smooth evolution of energies with increasing size.

A comparison of size-dependent total energy and energy density of $Ni_2Si$ and G-phase precipitates inside the Cu matrix obtained from DFT and predicted by M3GNet and 7net-Ompa-mpa models is shown in **Figure 5**. The plots of other **7net** models are shown in **Figure S1** and **S2**. Both MLIPs successfully reproduce the DFT-calculated energetic trends over the full particle-size range, including relative stability ordering between $Ni_2Si$ and G-phase. In particular, the MLIPs accurately capture the higher formation energy of G-phase at small cluster sizes and its gradual stabilization with increasing size relative to $Ni_2Si$. The curvature, crossover behavior, and asymptotic convergence of energy density at larger sizes are consistently reproduced, with only minor quantitative deviations observed near the smallest sizes. These results demonstrate that the employed MLIPs reliably replicate DFT predictions for competing precipitate phases inside the Cu matrix, validating their transferability across different crystal structures and size regimes.

### 3.5 Experimental observations of GB precipitates

**Figure 5** shows the grain boundary structure SEM images of Cu-Ni-Si and Cu-Ni-Si-Mn alloys obtained after homogenization and subsequent air cooling. The details of the alloy processing and experimental procedures are described in Ref-[24]. Homogeneous nucleation within grain interiors forms strained $Ni_2Si$ precipitates coherent with the Cu matrix, whereas heterogeneous nucleation at GBs produces much larger precipitates. As a result, a particle-free zone was formed around the GBs owing to the formation of GB particles. At GBs in Cu-Ni-Si alloys, irregularly shaped $Ni_2Si$ precipitates are observed, as shown in **Figure 5(a)**. In contrast, GBs in Cu-Ni-Si-Mn alloys exhibit film-shaped G-phase precipitates, as shown in **Figure 5(b)**, together with occasional spherical particles, consistent with previous reports-[24]. The size of $Ni_2Si$ at GBs is rather larger than that of $Mn_6Ni_{16}Si_7$ at GBs.

### 3.6 Strain-released interphase boundaries

The distinct GB precipitate morphologies of $Ni_2Si$ and $Mn_6Ni_{16}Si_7$ can be related to differences in interfacial energetics. To address this issue, we perform MLIP calculations of Cu/$Ni_2Si$ interfacial structures both for strained and strain-released structures using M3GNet model only.

A strained $Ni_2Si$ within the Cu grain interior is known to form a coherent interface with the Cu matrix: see **Figure 1(d)** for Cu$(1\bar{1}0)$/strained-$Ni_2Si(100)$ coherent interface structure. In M3GNet MLIP calculations, the lattice mismatches along $Ni_2Si$ [010] and $Ni_2Si$ [001] directions against coherent Cu lattices are calculated as -3.27% and 2.28%. The area of $Ni_2Si$ is different by -1.06% compared to coherent interface area of Cu. However, when $Ni_2Si$ is formed at GBs, it recovers original lattice parameters losing coherent relationship between Cu and $Ni_2Si$. Or, if $Ni_2Si$ precipitate size becomes larger, it may form a semi-coherent interface structure. To model this, we generate a partially-strain-released interphase boundary structure of Cu$(1\bar{1}0)$/$Ni_2Si(100)$, named **semi-IF1**, where $20 \times [010]_{Ni_2Si}$ and $24 \times [001]_{Ni_2Si}$ lattice vectors are strained to match $19 \times [1\bar{1}0]_{Cu}$ and $25 \times [001]_{Cu}$ lattice vectors: see **Figure 6**. The **semi-IF1** interface structure has a supercell size of 9.71 nm and 9.03 nm for in-plane supercell lattice vectors. And this supercell contains 11,520 Ni and Si atoms and 7,600

Cu atoms. Here, lattice strain on $Ni_2Si$ for these vectors are reduced to 1.82% and -1.82%, respectively. As one of two vectors is tensile strained while the others is compressed strained, the arial strain of $Ni_2Si$ in the **semi-IF1** structure is only -0.02%. **Figure 6(a)** shows the initial interface structure before the structural relaxation. We can find both in-phase coherent atomic stacking of $Ni_2Si$ and out-of-phase coherent atomic stacking of $Ni_2Si$ simultaneously, like semi-coherent structure. After structural relaxation using M3GNet MLIP calculations, we obtain the interface atomic structure as shown in **Figure 6(b)**. In this structure, we find no additional gap is needed between $Ni_2Si$ and Cu. Interestingly, in this supercell size, we observe that the chemical energy gain of the coherent interfacial bonding well overcomes the strain energy for $Ni_2Si$ bulk. As a result, we may distinguish a coherent zone with semi-coherent position where even bulk Ni and Si atoms are disordered.

**Figure 7(a) and (b)** show the artificially gapped interface structure for Cu$(1\bar{1}0)$/ $Ni_2Si(100)$, named **semi-IF1-1A,** before and after structural relaxation, respectively. The difference between **semi-IF1** and **semi-IF1-1A** is the additional gap of 1 Å between Cu and $Ni_2Si$ bulk. Again, coherent interface atomic stacking lowers the total energy of the system. However, as the distance between Cu and $Ni_2Si$ is increased in **semi-IF1-1A** compared to **semi-IF1**, the formation of coherent interfacial bonding causes an additional bulk strain along the normal direction to the interface. As a result, the area size of attractive in-plane coherent region is reduced, while the area of out-of-phase repulsive zone is increased. The strain on the repulsive zone $Ni_2Si$ is decreased and the disordered behavior is also reduced. When the size of the additional gap is increased to 1.2 Å, where the interface structure is named as **semi-IF1-1.2A**, the size of in-plane stacking area is further reduced while the out-of-plane stacking area is increased. Note that the in-plane zone shows an attractive interaction between interfacial $Ni_2Si$ and Cu, while the out-of-plane zone shows a repulsive interaction with an open-boundary interface structure.

**Figure 8** shows the in-phase and out-of-phase coherent interface structure for Cu$(1\bar{1}0)$/ $Ni_2Si(100)$. As reported, $Ni_2Si$ can have a perfectly matched coherent relationship with Cu. However, at GBs, the strain released structure will have a phase shift in $Ni_2Si$ atomic positions with respect to the Cu, even for coherent interface for Cu$(1\bar{1}0)$/ $Ni_2Si(100)$. If it happens, it will cause repulsive interaction as shown in **Figure 8(b).** We find that the optimal gap between Cu and $Ni_2Si$ is computed as 0.66 Å for out-of-phase structure, while it is zero for in-phase structure. This result highlights the possible formation of open-boundary structure after stress release via formation of the out-of-phase zone.

**Table 2** summarizes the predicted energies of various interface and surface atomic structure configurations between Cu and $Ni_2Si$ using M3GNet MLIP calculations. Here we do not include bulk strain release effect as we extract each energetic contribution of bulk, strain, and interface energies. The in-phase coherent interface structure shows the lowest energy among considered structures: 0.311 $J/m^2$. For semi-coherent interface structure, the interface energy is 0.427 $J/m^2$, higher than that of coherent owing to partial loss of coherency. The incoherent interface energy is quite larger than coherent and semi-coherent ones: 0.541-650 $J/m^2$, depending on configurations. These values are quite lower than the surface energy of 1.07 $J/m^2$. However, the inclusion of out-of-phase zone cause a high interface energy indicating a repulsive interaction between Cu and $Ni_2Si$. It is highlighted in the out-of-phase coherent interface structure with its very high interface energy of 1.199 $J/m^2$. Note that this value is higher than the surface energy. Overall, we find directional and phase dependent interface energies in Cu/$Ni_2Si$ interface structure for Cu-Ni-Si alloys.

## 4. Discussion

### 4.1 Mechanisms governing precipitate phase selection

The relatively higher thermodynamic stability of G-phase upon the addition of Mn to the solid solution suggests that Mn plays a key role in promoting the thermodynamic preference for G-phase over competing precipitate phases. **Figure 2** presents the formation energies of various Ni-Si-Mn solute configurations in the Cu solid solution, referenced to the $\delta$-$Ni_2Si$ + $\alpha$-Mn state. This reference choice allows a direct assessment of the thermodynamic tendency for phase transformation involving $Ni_2Si$ and Mn. Among the $Ni_2Si$-containing configurations, the addition of Mn generally increases the formation energy, indicating that isolated $Ni_2Si$ precipitates coexisting with Mn are not the most stable state once sufficient Mn is available. In contrast, configurations corresponding to the G-phase exhibit negative formation energies relative to the reference, particularly for solute Si-rich states. This clearly demonstrates that the G-phase is thermodynamically favored over $\delta$-$Ni_2Si$ + $\alpha$-Mn when Ni, Mn, and Si are allowed to rearrange cooperatively within the Cu matrix. Although Ni$_2$Si forms readily at early stages, the formation energy analysis shows that continued Mn enrichment provides a strong thermodynamic driving force for transformation into the G-phase. The excess Si released during this process can remain in solid solution or segregate locally, further stabilizing the G-phase configuration. Therefore, $Ni_2Si$ should be viewed as a metastable precursor that forms first but progressively transforms into the G-phase as diffusion proceeds and local chemical equilibrium is approached. This explains why $Ni_2Si$ is often observed at early aging stages, while the G-phase emerges as the dominant phase at later stages, particularly in regions where solute redistribution is kinetically accessible, such as grain boundaries.

The enhanced stability can be understood in terms of the ability of G-phase to accommodate Mn within its complex crystal structure, thereby reducing the overall energetic penalty associated with phase formation in the Cu matrix. In contrast, simpler intermetallic phases exhibit higher relative energies, reflecting less favorable accommodation of Mn. Taken together, these results support the conclusion that Mn addition shifts the energetic balance toward G-phase formation, making it the most stable precipitate under the investigated conditions. It should be emphasized that these energies reflect bulk chemical stability and do not account for elastic strain, surface or interfacial contributions associated with precipitation. Consequently, while the present analysis establishes an intrinsic thermodynamic preference for the G-phase at the compositional level, the actual precipitation behavior must be further assessed by incorporating size-dependent strain and interfacial effects, as discussed in the following sections.

### 4.2 Origin of $Ni_2Si$ precipitate within the grain interior

The energy density and total formation energy of plate-like $Ni_2Si$ within the Cu matrix and at the grain boundary are shown in **Figure 4**. Here, we have compared the DFT results only with M3GNet and 7net-Ompa-mpa models across a wide range of precipitate sizes. The data of other MLIP models can be found in the supplementary information. At the early stages, there is a low possibility of precipitate formation both inside the matrix and outside the grain due to the positive energy density. The minimum particle size for nucleation to occur is about 0.2 nm inside the matrix and 0.6 nm outside it. Within this size regime, the particle observes a nucleation energy barrier on the total energy scale, which prevents precipitate growth. The barrier is small inside the matrix but more significant at the grain boundary, limiting nucleation outside the Cu matrix. As a result, $Ni_2Si$ initially begins to grow inside the matrix, and its energy density and total formation energy decrease steadily as the size increases. The semi-coherent nature of the $Ni_2Si$/Cu interface within the matrix allows elastic strain to spread over a larger volume which lowers the overall strain energy density as the precipitate grows. This energetic balance strongly favors intra-granular nucleation and growth, especially in plate-like shapes that minimize strain along elastically soft directions while maintaining semi-coherent interfaces.

At sizes exceeding 10 nm, a crossover of energy density of $Ni_2Si$ inside and outside the matrix is observed and approaches the saturated values. This value represents the bulk limit, where surface and interface contributions become insignificant, leaving only the strain energy term. Therefore, compared to the inside grain, $Ni_2Si$ has lower energy at the grain boundary, equal to the strain energy inside the matrix. This implies that $Ni_2Si$ may be more stable at the grain boundary after reaching a critical size, but the nucleation barrier at early stages restricts its formation. Thus, even with generally increased diffusivity at grain boundaries, the current energetics indicate that grain boundaries do not provide a sufficiently low-energy environment to offset the combined interfacial and strain penalties associated with $Ni_2Si$ formation. These results offer a thermodynamic foundation for the experimentally observed intra-granular, plate-shaped $Ni_2Si$ precipitates in Cu-based alloys.

### 4.3 Competing stability of $Ni_2Si$ and G-phase at grain boundary

The size dependence of the total energy and energy density for intragranular $Ni_2Si$ and G-phase precipitates embedded in the Cu matrix offers direct insight into the thermodynamic factors that control phase selection during precipitation. Experimental results indicate that adding Mn to the system impacts both the formation and growth of $Ni_2Si$ precipitate inside and outside the Cu matrix. It has been established that plate-shaped $Ni_2Si$ forms and stabilizes quickly within the matrix compared to at the grain boundary. To examine the particle nucleation process and the total formation energy of the G-phase within a Cu matrix, we compared it with the nucleation and total formation energy of $Ni_2Si$. **Figure 9** shows that at small precipitate sizes, the G-phase has a significantly higher nucleation barrier than $Ni_2Si$, suggesting that its formation within the Cu matrix is not favorable during the early stages of precipitation. This behavior mainly stems from the much higher interfacial energy related to the Cu/G-phase interface, which dominates the total free energy at small sizes. In contrast, $Ni_2Si$ has a negligible nucleation barrier, even though it has a relatively higher bulk formation energy. This occurs because its lower interfacial energy greatly reduces the surface contribution to the nucleation free energy. As a result, $Ni_2Si$ can easily nucleate within the grain interior, while G-phase nucleation faces both kinetic and thermodynamic challenges at smaller sizes. As precipitate size increases, the bulk energy term gradually outweighs the interfacial contribution for both phases, leading to a steady decrease in total energy. In this range, the energy difference between $Ni_2Si$ and G-phase reduces, and both phases approach their bulk thermodynamic limits, including a minor contribution from elastic strain energy. These findings highlight the significant role of interfacial energy during the early stages of precipitation. At small precipitate sizes, the total energy is primarily influenced by the interface with the Cu matrix rather than by the bulk of the phase. As the precipitates grow larger, the bulk energy contribution becomes more important and eventually dominates the total energy. This explains why the G-phase stabilizes inside the Cu matrix only at larger sizes or under conditions that lower interfacial energy.

The energy density trends in the second row of **Figure 9** highlight the competitive stability of $Ni_2Si$ and the G-phase within the Cu matrix. For both phases, the energy density decreases steadily as the precipitate size increases. This reflects a gradual drop in interfacial contributions, while bulk and strain-related factors become more significant. Importantly, a critical size is identified where the energy density of the G-phase falls below that of $Ni_2Si$. This indicates that the G-phase only becomes energetically favorable beyond this threshold size. Below this critical size, the G-phase remains unfavourable because of its significantly higher interfacial energy, which creates a substantial nucleation barrier. After crossing the size threshold, the influence of interface energy becomes minimal. The relative stability of the two phases is mainly determined by elastic strain energy. Since the G-phase has lower strain energy than $Ni_2Si$, it becomes more stable thermodynamically at larger sizes. However, at smaller sizes, the G-phase faces a significantly higher nucleation barrier, which greatly restricts its formation within the Cu matrix. Therefore, although the G-phase shows greater stability at larger sizes, its formation within the matrix is hindered by this nucleation barrier. It tends to form preferentially at grain boundaries, where both interfacial and strain energy penalties are lower.

Based on the findings in **Table 3**, the DFT analysis indicates that the nucleation barrier for precipitation of $Ni_2Si$ is relatively higher (~ 1.23 eV) outside of the Cu matrix, while it decreases dramatically to ~ 0.01 eV within the matrix. Additionally, the critical size needed to surpass the nucleation barrier for $Ni_2Si$ precipitate within the Cu matrix (~ 0.15 nm) is significantly smaller than that at the grain boundary (~ 0.60 nm). This clearly demonstrates that the Cu matrix greatly promotes the nucleation of $Ni_2Si$ by considerably lowering the barrier. The elevated barrier outside the matrix suggests a delayed thermodynamic stabilization of $Ni_2Si$ clusters at grain boundaries, whereas phase stability is enhanced within the Cu matrix. Quantitatively, DFT results show a notable difference between the nucleation behaviors of $Ni_2Si$ and the G-phase inside the Cu matrix. The critical formation size for $Ni_2Si$ (~ 0.15 nm) is more than four times smaller than that of the G-phase (~ 0.70 nm), indicating that the G-phase experiences a much more delayed nucleation process within the matrix. Consistently, the nucleation barrier for the G-phase is considerably higher (~ 2.45 eV) compared to the mere ~ 0.01 eV for $Ni_2Si$, pointing to significant differences in nucleation kinetics. As particle sizes increase, the total energy of the G-phase decreases much more steeply than that of $Ni_2Si$, making it thermodynamically more advantageous at a size of approximately 2.4 nm in comparison to $Ni_2Si$. Ultimately, the combination of large critical formation size, an elevated nucleation barrier, and relatively high critical crossing size renders the G-phase thermodynamically and kinetically less favorable for nucleation within the Cu matrix. Although it becomes energetically competitive with $Ni_2Si$ at larger cluster sizes, the significantly greater nucleation barrier of the G-phase inhibits its early formation within the bulk material. Consequently, nucleation is preferentially directed to grain boundaries. Together, these elements provide a clear thermodynamic explanation for why the G-phase is regularly found at grain boundaries rather than in the Cu matrix, aligning well with both DFT predictions and MLIP-based observations.

All MLIPs successfully capture this pronounced difference between formation and crossing behavior as shown in **Figure S1 and S2**. In particular, M3GNet and the 7net variants reasonably reproduce the large crossing size inside Cu, preserving both the magnitude and physical trends observed in the DFT calculations. This close agreement demonstrates that MLIPs not only reproduce DFT-level critical sizes quantitatively but also reliably capture the underlying phase competition within the Cu matrix.

### 4.4 Incoherent interface energies and GB precipitate morphologies

The distinct GB precipitate morphologies of $Ni_2Si$ and $Mn_6Ni_{16}Si_7$ in Cu alloys can be understood in terms of differences in interfacial energetics as well as coherent versus incoherent bonding character. For $Ni_2Si$, the Cu/$Ni_2Si$ interface exhibits both low- and high-energy configurations. Even when a coherent interface is initially formed, strain relief during precipitate growth can induce the formation of an out-of-phase coherent-like structure due to lattice mismatch, creating a locally repulsive interfacial region. The energy of such an out-of-phase coherent-like interface can exceed that of a fully incoherent interface. As a result, the interfacial energy of $Ni_2Si$ precipitates varies strongly with interface orientation and atomic configuration. Our MLIP calculations thus provide an atomistic explanation for the irregularly shaped, strain-relieved $Ni_2Si$ precipitates observed at grain boundaries. In contrast, the film-shaped $Mn_6Ni_{16}Si_7$ precipitate, spanning adjacent Cu grains, suggests relatively weak orientation dependence of the interfacial energy. Because the two neighboring Cu grains have different crystallographic orientations, the two sides of a film-shaped G-phase precipitate must form distinct incoherent interfaces with the Cu matrix, as reported in Ref-[24]. Nevertheless, the existence of the film-shaped morphology indicates that these interfaces remain comparatively stable despite their structural differences, as predicted in Ref-[24].

Overall, the orientation and configuration dependence of interfacial energetics appears to play a key role in determining GB precipitate morphology in Cu-Ni-Si-based alloys, thereby contributing to mechanical-property enhancement via heterogeneous nucleation of G-phase with Mn addition and degradation in Mn-free alloys after heterogeneous nucleation of irregular shaped $Ni_2Si$ at GBs.

## 5. Conclusion

This study demonstrates that modern machine-learning interatomic potentials based on the M3GNet and 7net frameworks can reliably reproduce DFT-level energetics for complex Cu, Ni, Si, and Mn alloy systems across multiple length scales. The main findings of this study can be summarized as follows:

(1) Formation energy calculations confirm that incorporating Mn helps stabilize Mn–Ni–Si (G-phase type) configurations in the Cu solid solution. This provides a thermodynamic driving force for transforming $Ni_2Si$-related solutes toward the G-phase.

(2) By breaking down bulk, surface, interfacial, and strain energy contributions, we clarify how size affects the stability of $Ni_2Si$ and G-phase precipitates within the Cu matrix.

(3) At small precipitate sizes, $Ni_2Si$ is favored due to lower interfacial energy within the Cu, leading to a small nucleation barrier and facilitating its formation within the grains. Although, the elastic strain energy of G-phase is lower, its higher interfacial energy results in a significant nucleation barrier at small sizes.

(4) As precipitate size increases, interfacial effects become less important, and bulk energy takes over, leading to stability crossover in which the G-phase becomes thermodynamically favored. However, a large nucleation barrier limits its formation inside the matrix. This explains its preferential formation at grain boundaries.

(5) All qualitative trends and quantitative energy differences obtained from DFT are consistently reproduced by the MLIP models. This includes solid-solution energetics, size-dependent stability, and $Ni_2Si$ to G-phase crossover behavior.

(6) The morphology of grain boundary precipitates in Cu-Ni-Si alloys is governed by the orientation and configuration dependence of interfacial energetics. $Ni_2Si$ forms irregular shapes due to high-energy, strain-induced repulsive regions, while the G-phase adopts a film-like structure supported by stable, orientation-independent incoherent interfaces. This transition explains why Mn additions enhance mechanical properties by favoring stable G-phase nucleation over degraded, irregular $Ni_2Si$ formations.

Importantly, these results demonstrate that MLIPs are strong and efficient tools for capturing complex precipitation energetics and microstructural mechanisms. They enable predictive modeling of phase selection in engineering alloys beyond the practical limits of traditional first-principles calculations.

**CRediT authorship contribution statement**

**Aadil Fayaz Wani:** Writing – original draft, Conceptualization, Data curation, Investigation, Visualization, Writing - review and editing. **Il-Seok Jeong:** Validation, Formal analysis, Writing - review and editing. **Haekwan Jeon:** Validation, Formal analysis, Writing - review and editing. **Jaesun Kim.** Validation, Formal analysis, Writing – review and editing. **SuDong Park:** Conceptualization, Funding acquisition, Writing - review and editing. **Eun-Ae Choi:** Conceptualization, Formal analysis, Validation, Resources, Project administration. **Seung Zeon Han:** Conceptualization, Formal analysis, Funding acquisition, Resources, Project administration. **Seungwu Han:** Formal analysis, Validation, Funding acquisition, Resources, Project administration. **Byungki Ryu:** Conceptualization, Data curation, Investigation, Formal analysis, Funding acquisition, Resources, Project administration, Writing - review and editing.

**Declaration of competing interest**

The authors declare that they have no known competing financial interests or personal relationships that could have appeared to influence the work reported in this paper

**Acknowledgments**

This work was supported by the National Research Foundation of Korea (NRF) grant funded by the Korean Government (MSIT) (Grant No. 2022M3C1C8093916, which has been assigned a new number, **RS-2022-NR119739**). It is also was supported by the Primary Research Program of KERI through the National Research Council of Science and Technology (NST) funded by the Ministry of Science and ICT (MSIT) (Grant No. **26A01005**) and by the Korea Institute of Energy Technology Evaluation and Planning (KETEP) grant funded by the Ministry of Trade, Industry and Energy (MOTIE) (Grant No. 2021202080023D, which has been assigned a new number, **RS-2021-KP002416**). It was also partially supported by a National Research Foundation of Korea (NRF) grant funded by the Korean Government (MSIT) (Grant No. **RS-2025-25461730**).

# Tables and Figures

**Table 1.** Heat of formation for δ-$Ni_2Si$ and G-phase calculated using DFT and MLIP calculations. Percentage errors are evaluated relative to DFT values.

| Calculation method | | δ-$Ni_2Si$ | | $Mn_6Ni_{16}Si_7$ | |
|---|---|---|---|---|---|
| | | $E_f$ (eV/atom) | Error in $E_f$ | $E_f$ (eV/atom) | Error in $E_f$ |
| | DFT | -0.482 | 0% | -0.616 | 0% |
| MLIP | M3GNet | -0.559 | -16.0% | -0.687 | -11.520 |
| | 7net-0 | -0.560 | -16.1% | -0.676 | -9.740 |
| | 7net-Ompa-mpa | -0.542 | -12.4% | -0.670 | -8.844 |
| | 7net-Ompa-omat | -0.544 | -12.9% | -0.634 | -2.900 |
| | 7net-Omni-mpa | -0.555 | -15.145 | -0.660 | -7.143 |
| | 7net-Omni-omat | -0.552 | -14.523 | -0.628 | -1.948 |

**Table 2**. Predicted interphase interface energies using M3GNet MLIP calculations for Cu/$Ni_2Si$ interface and $Ni_2Si$ surface structures

| Interface type (number of atoms) | Structure name | Boundary Energy ($J/m^2$) | $Ni_2Si$ plane | Cu plane | Spacing (Å) |
|---|---|---|---|---|---|
| **Coherent (48 atoms)** | In-phase<br>Out-of-phase | 0.311<br>1.199 | (100)<br>$1 \times [010]$,<br>$1 \times [001]$ | $(\bar{1}10)$<br>$1 \times [110]$,<br>$1 \times [001]$ | 0.00 (opt)<br>0.66 (opt) |
| **Semi-coherent (19,120 atoms)** | semi-IF1<br>semi-IF1-1Å<br>semi-IF1-1.2Å | 0.427<br>1.027<br>1.258 | (100)<br>$20 \times [010]$,<br>$24 \times [001]$ | $(\bar{1}10)$<br>$19 \times [110]$,<br>$25 \times [001]$ | 0.00 (opt)<br>1.0 (fixed)<br>1.2 (fixed) |
| **Incoherent (96 atoms)** | IF3 | 0.541 | (010)<br>$1 \times [100]$,<br>$1 \times [010]$ | (010)<br>$2 \times [100]$,<br>$1 \times [010]$ | 0.2 (opt) |
| **Incoherent (960 atoms)** | IF5 (twist Σ5 of IF3) | 0.650 | (010)<br>Σ5 of IF3 | (010)<br>Σ5 of IF3 | 0.2 (fixed) |
| **Surface** | | 1.07 | (100) | n/a | Vacuum |

**Table 3.** Critical sizes for nucleation and energetic crossover of $Ni_2Si$ and G-phase precipitates inside and outside the Cu matrix, obtained from DFT and various machine learning interatomic potential (MLIP) models.

| **Model** | **$Ni_2Si$** | | | **G-phase** | |
|---|---|---|---|---|---|
| | **Critical size of formation, $d_c$** | | **Critical size of crossing** | **Critical size of formation, $d_c$** | **Critical size of crossing** |
| | **Outside** | **Inside** | **Inside** | **Inside** | **Inside** |
| DFT | 0.6 | 0.15 | 35.1 | 0.7 | 2.4 |
| M3GNet | 0.3 | 0.10 | 11.8 | 0.5 | 2.2 |
| 7net-0 | 0.4 | 0.20 | 14.1 | 0.7 | 5.4 |
| 7net-Ompa-mpa | 0.5 | 0.15 | 20.3 | 0.7 | 5.4 |
| 7net-Ompa-omat | 0.6 | 0.15 | 20.3 | 0.6 | 5.4 |
| 7net-Omni-mpa | 0.6 | 0.15 | 20.3 | 0.6 | 5.4 |
| 7net-Omni-omat | 0.6 | 0.15 | 20.3 | 0.6 | 5.4 |

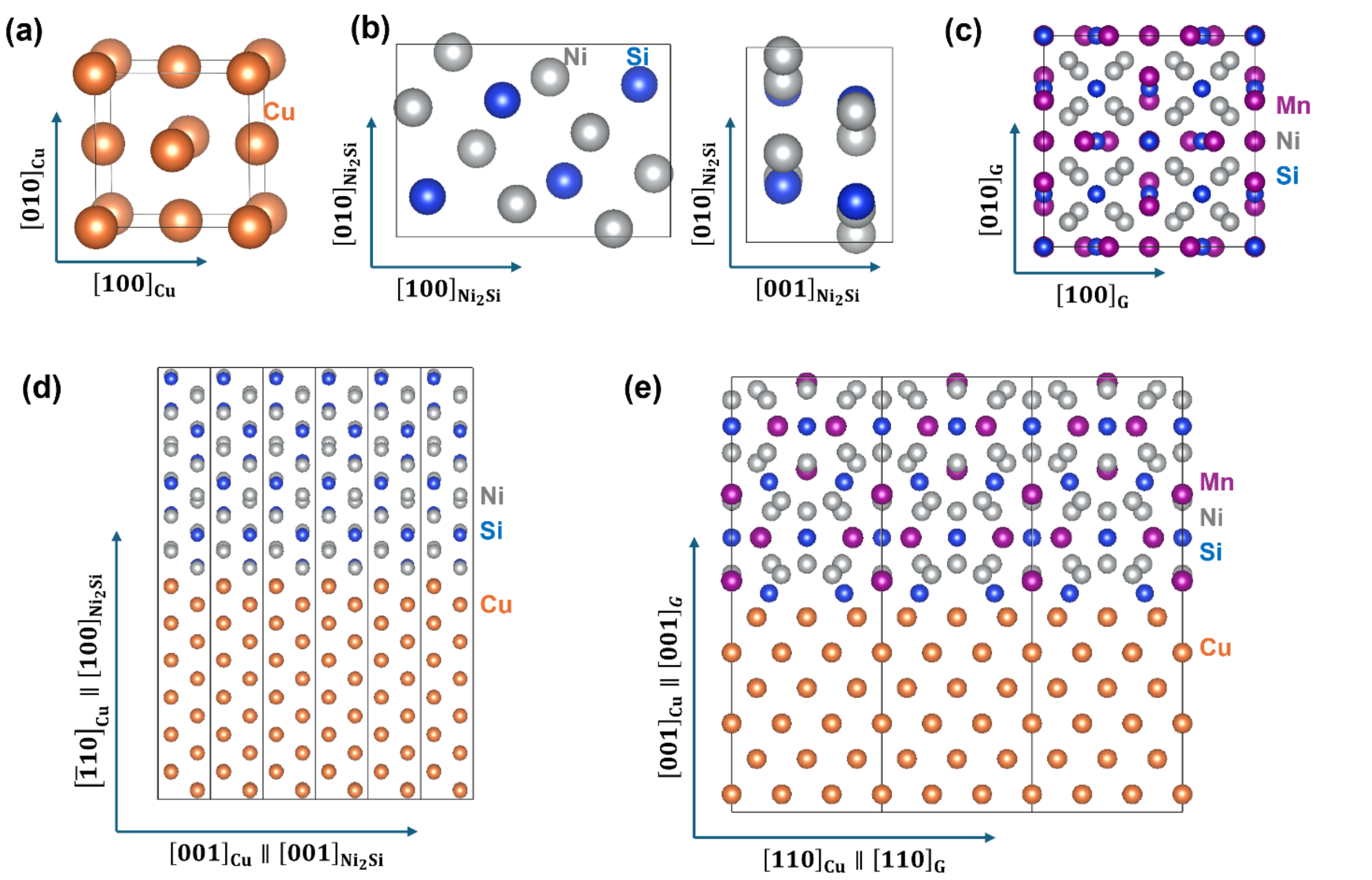

**Figure 1.** Atomic structure (a) Cu, (b) $Ni_2Si$, and (c) G-phase. The crystallographic directions are shown in each structure. correspond to bulk, bulk strained, and bulk strained G-phase respectively. The interface configurations in (d) and (e) correspond to the interface structures for Cu($1\bar{1}0$)/$Ni_2Si$(100) and Cu(100)/G-phase(100), respectively.

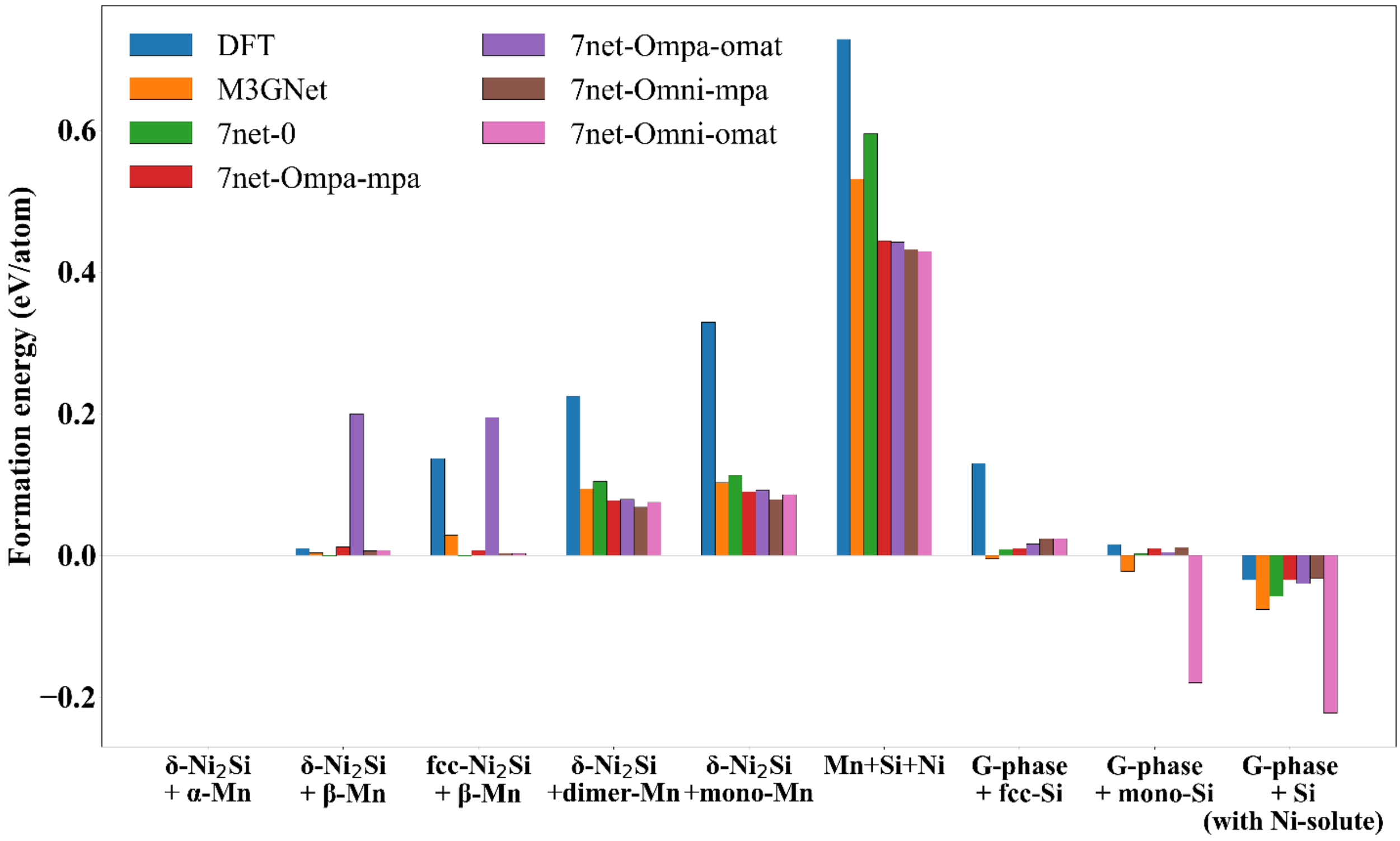

**Figure 2.** Formation energy evolution of Cu solid solution with different Ni, Si, and Mn solute states. The values are obtained from DFT and various MLIPs for direct comparison of energetic trends and relative accuracy.

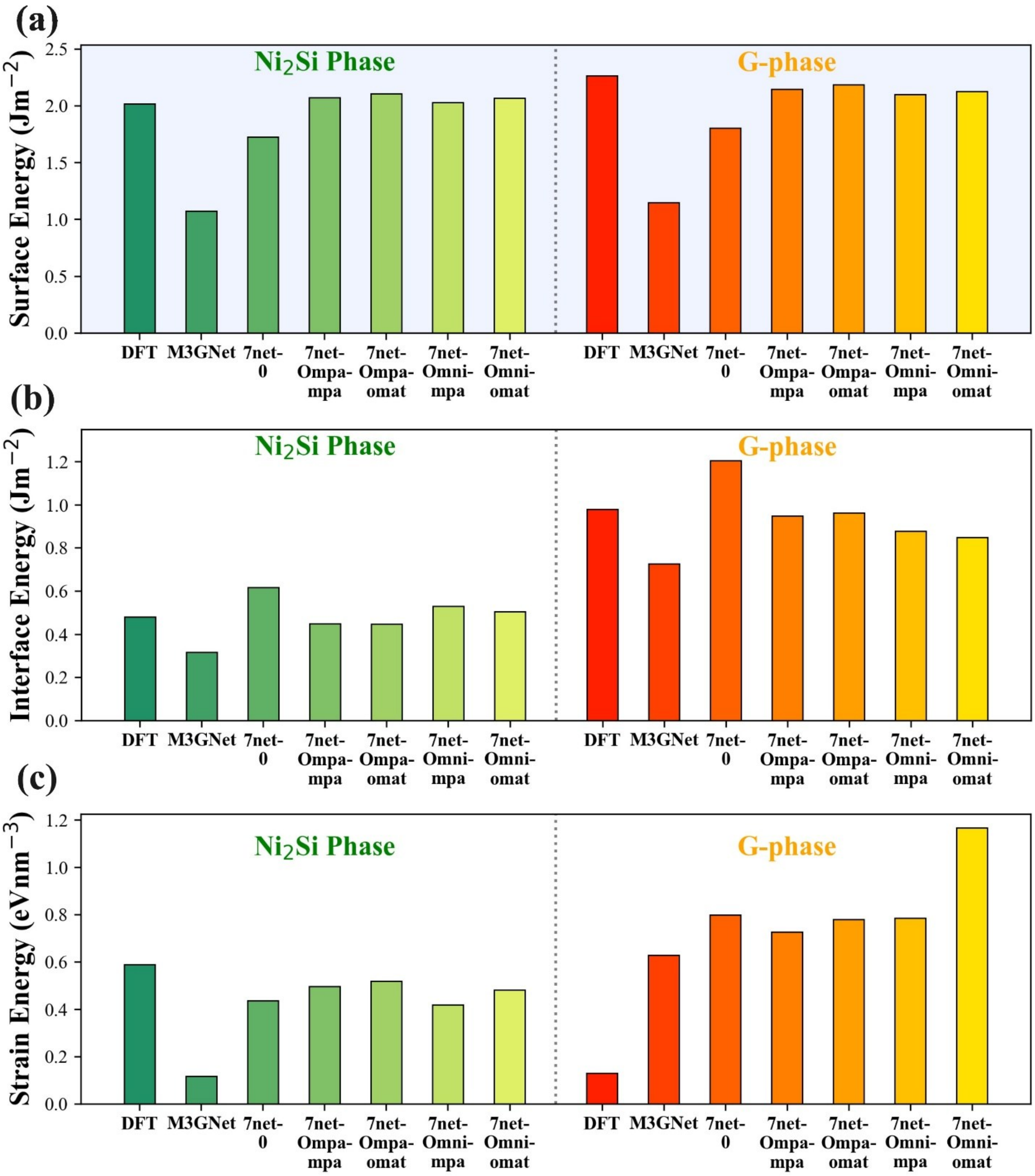

**Figure 3.** Comparison of (a) surface energy, (b) interface energy and (c) strain energy for $Ni_2Si$ and G-phase obtained from DFT and various MLIPs

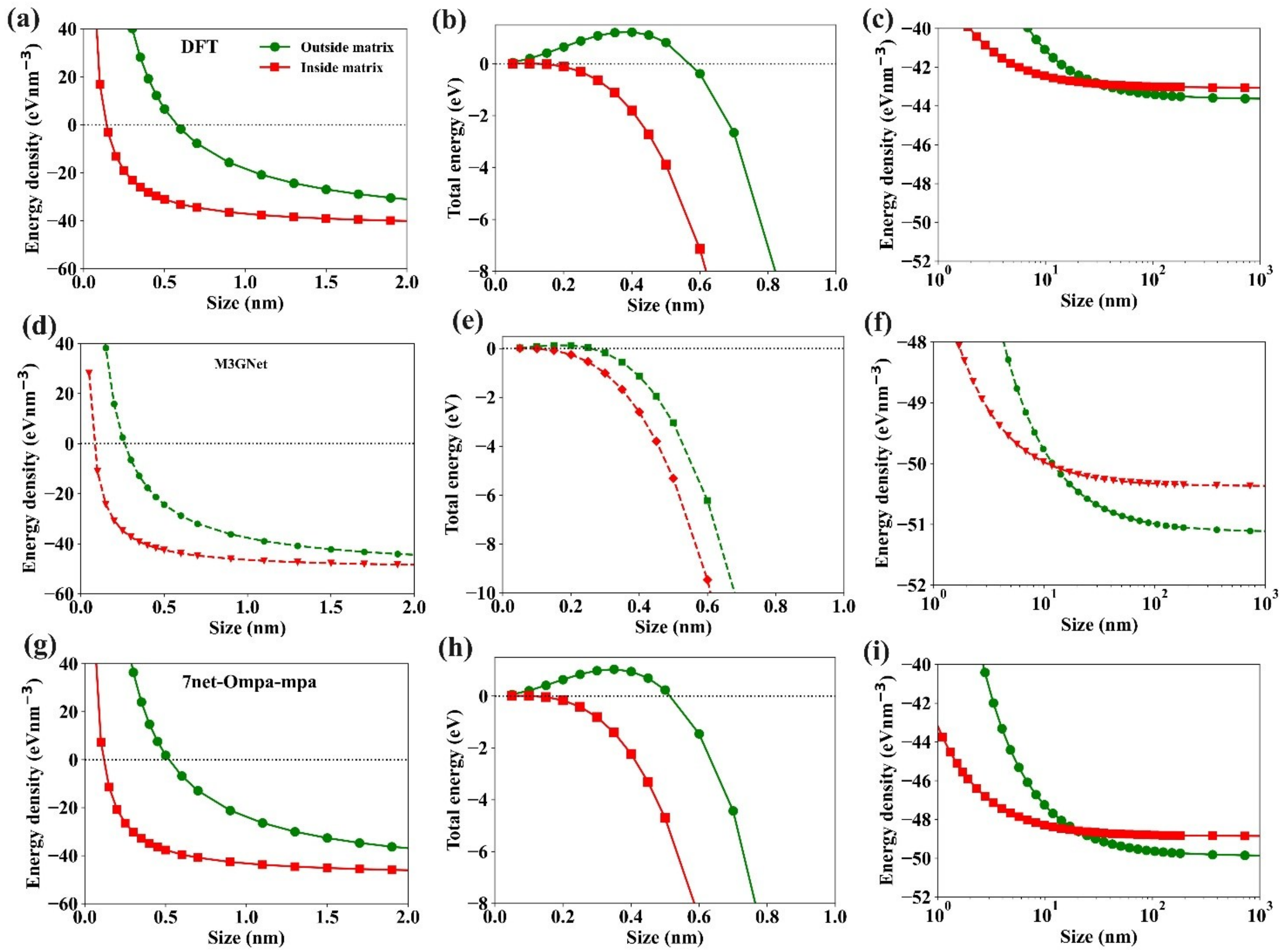

**Figure 4.** Comparison of (a, d, g) energy density at smaller particle sizes, (b, e, h) total energy at smaller sizes, and (c, f, i) energy density at larger sizes of $Ni_2Si$ precipitates inside and outside the Cu matrix computed using DFT and MLIP models.

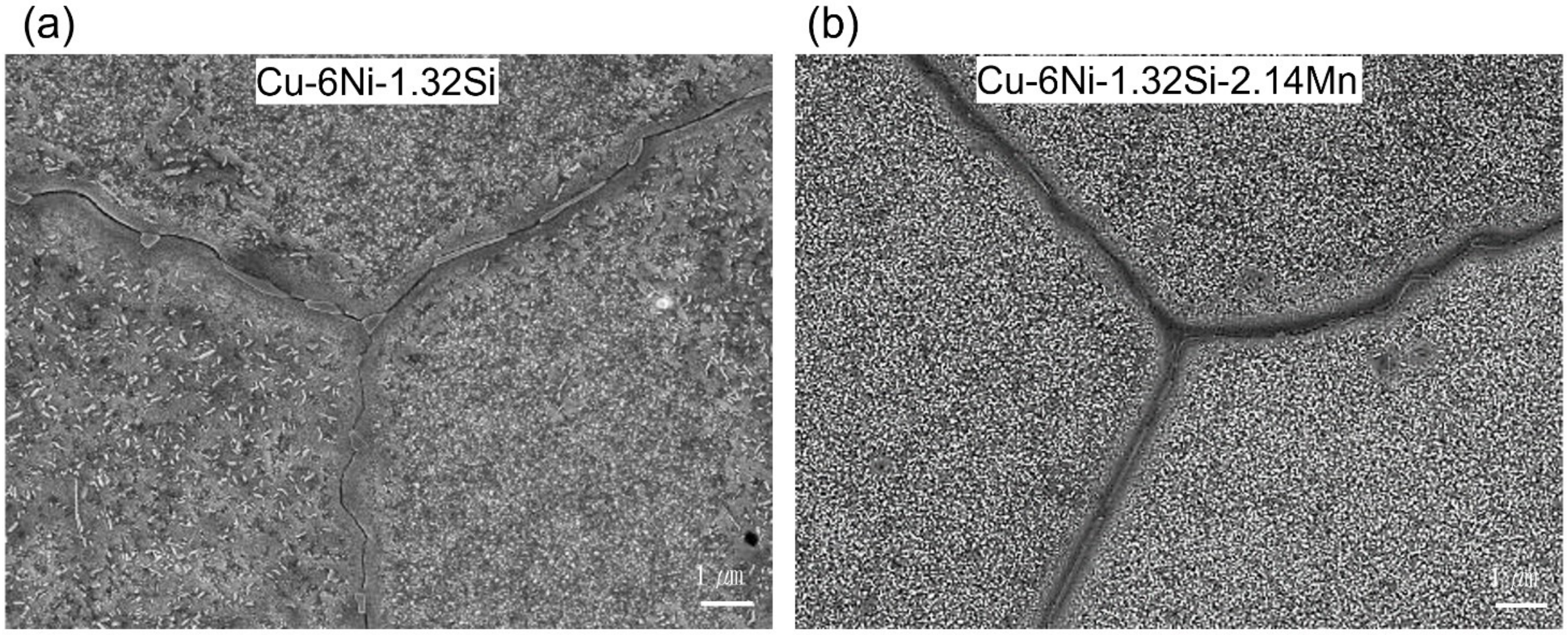

**Figure 5.** (a) SEM images around triple-junction grain boundary structures for Cu-Ni-Si and (b) Cu-Ni-Si-Mn alloys.

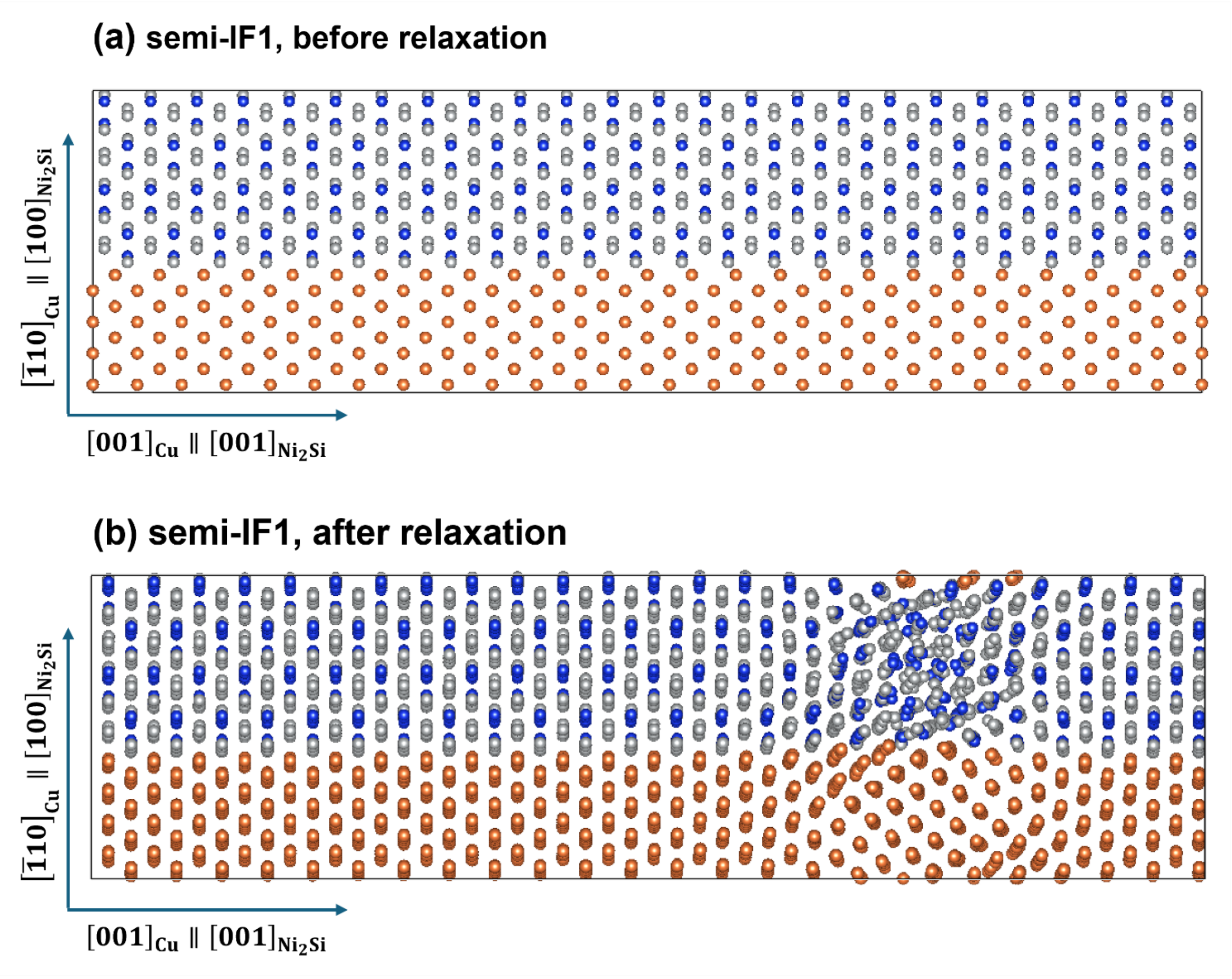

**Figure 6.** Atomic structure of partially-strain-released interface structure of **semi-IF1** for Cu/$Ni_2Si$ (a) before and (b) after structural optimization. Structural optimizations are performed using M3GNet MLIP calculations.

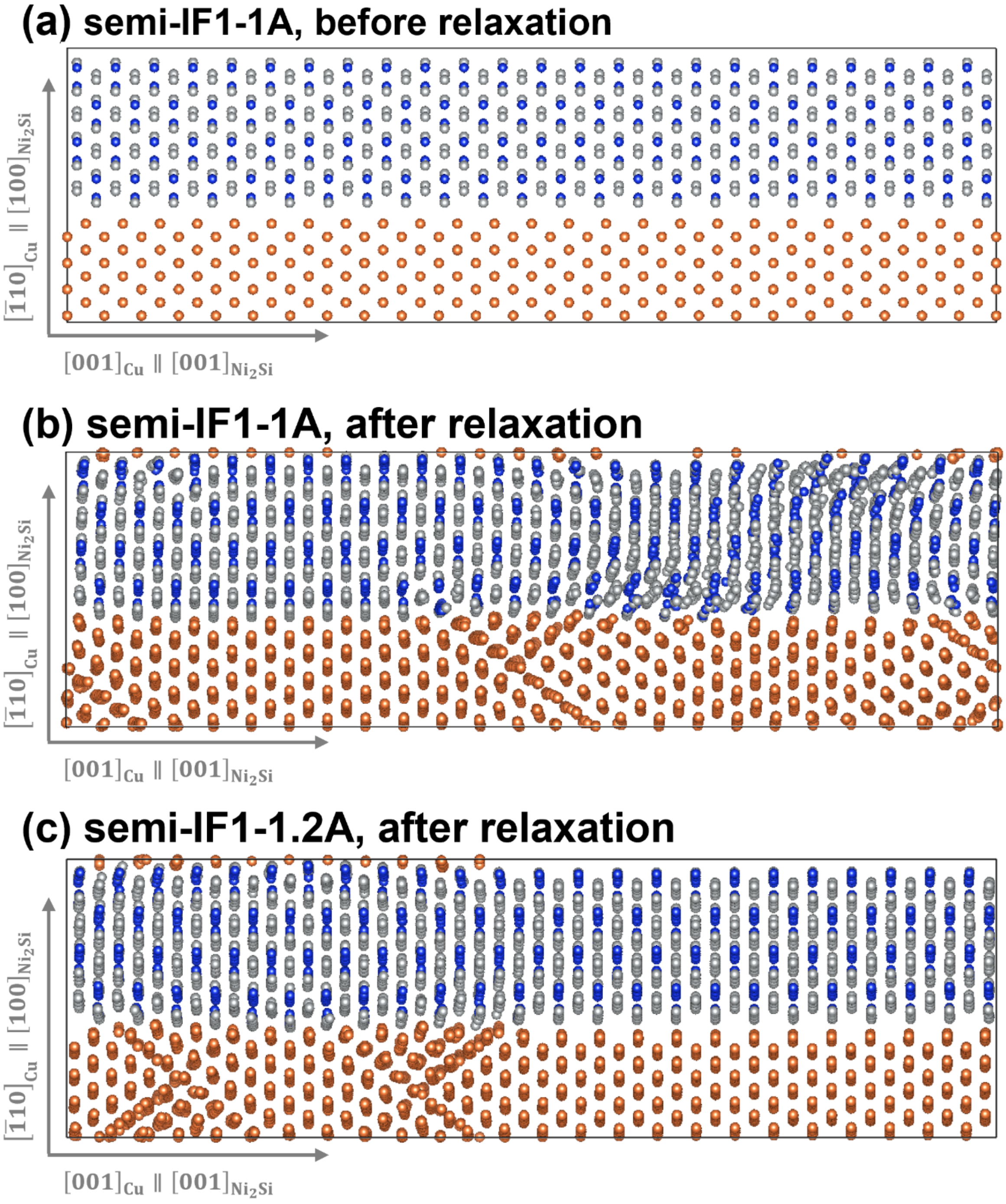

**Figure 7.** Atomic structure of partially-strain-released interface structure of **semi-IF1-1A** for Cu/$Ni_2Si$ (a) before and (b) after structural optimization, where the additional gap of 1 Å is inserted between Cu and $Ni_2Si$ artificially compared to **semi-IF1**. The atomic structure of **semi-IF1-1.2A** is shown in (c), where the additional gap of 1.2 Å is inserted between Cu and $Ni_2Si$. Structural optimizations are performed using M3GNet MLIP calculations.

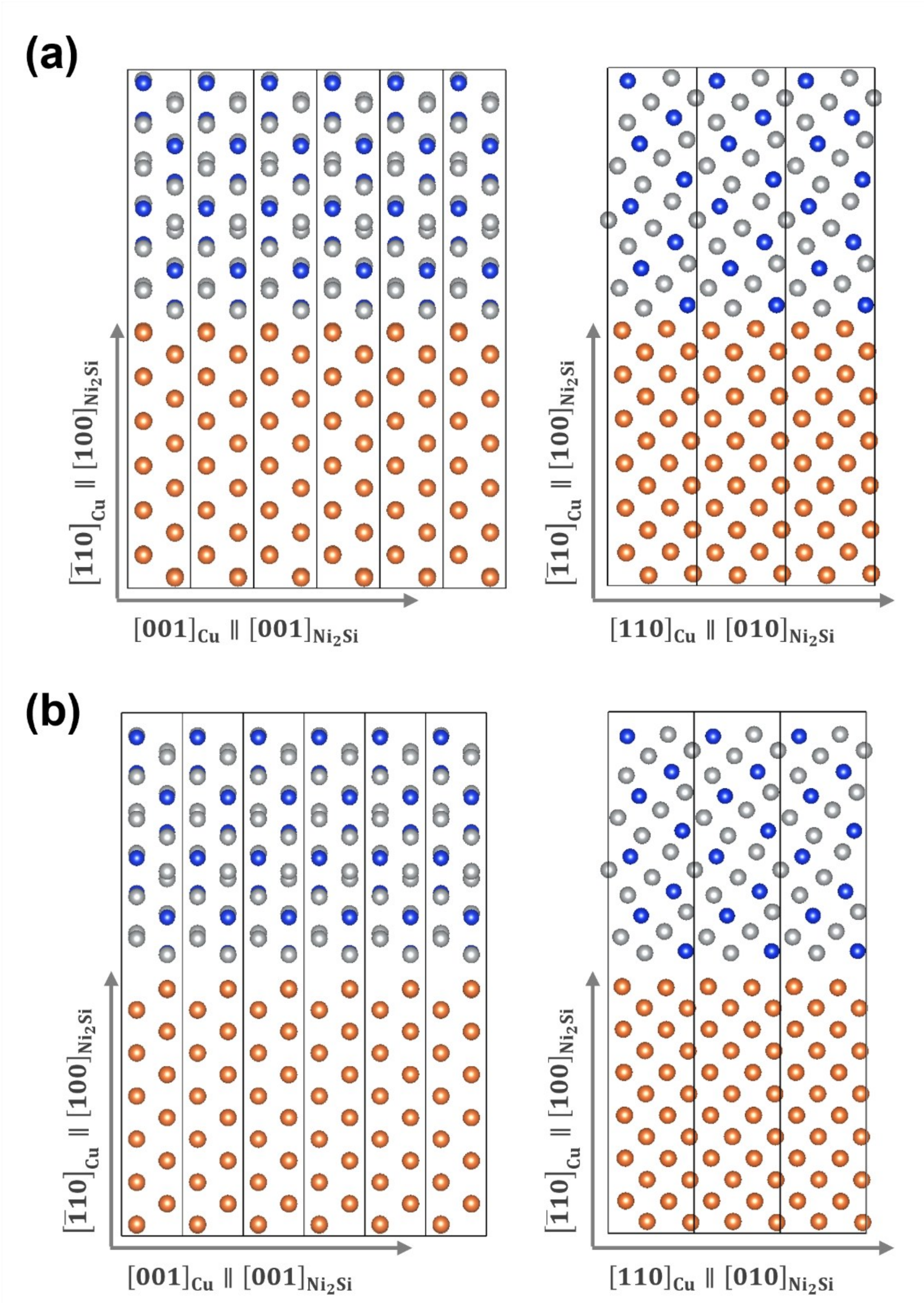

**Figure 8.** (a) In-phase and (b) out-of-phase coherent interface structure of Cu$(1\bar{1}0)$/strained-$Ni_2Si$(100). Structural optimizations are performed using M3GNet MLIP calculations.

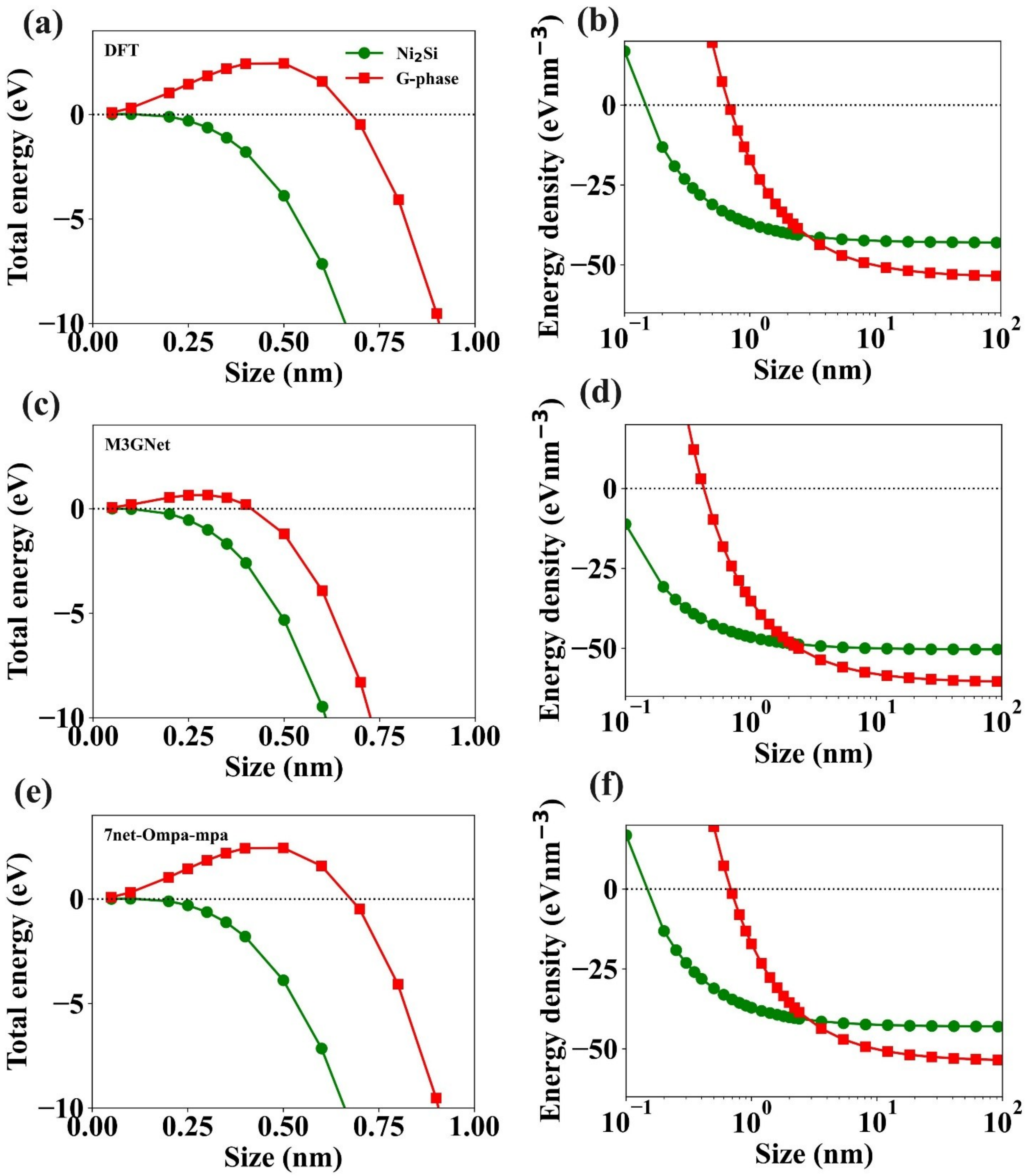

**Figure 9.** Comparison of size-dependent (a-c) total energy and (d-f) energy density of $Ni_2Si$ and G-phase computed using DFT and MLIP models

**References**

[1] S.Z. Han, E.-A. Choi, S.H. Lim, S. Kim, J. Lee, Alloy design strategies to increase strength and its trade-offs together, Prog. Mater. Sci. 117 (2021) 100720. https://doi.org/10.1016/j.pmatsci.2020.100720.

[2] L. Feng, J. Li, Q. Lu, Y. You, Z. Xu, L. Liu, L. Fu, P. Gao, J. Yi, C. Li, Accelerated development of high-strength and high-conductivity Cu-Cr-Ti alloys based on data-driven design and experimental validation, Mater. Des. 253 (2025) 113948. https://doi.org/10.1016/j.matdes.2025.113948.

[3] J. Huang, Z. Xiao, J. Dai, Z. Li, H. Jiang, W. Wang, X. Zhang, Microstructure and Properties of a Novel Cu–Ni–Co–Si–Mg Alloy with Super-high Strength and Conductivity, Mater. Sci. Eng. A 744 (2019) 754–763. https://doi.org/10.1016/j.msea.2018.12.075.

[4] N. Guo, Q. Cheng, X. Zhang, Y. Fu, L. Huang, Microstructure and Mechanical Properties of Underwater Laser Welding of Titanium Alloy, Materials 12 (2019) 2703. https://doi.org/10.3390/ma12172703.

[5] H. Yang, Y. Fan, J. Zhou, B. Shi, J. Chen, N. Lin, C. Wang, L. Xiao, Simultaneous enhancement of mechanical properties and electrical conductivity in Cu-Ni-Si alloy by constrained groove pressing and aging treatments, Mater. Sci. Eng. A 938 (2025) 148465. https://doi.org/10.1016/j.msea.2025.148465.

[6] J. Kang, H. Tan, Q. Lei, Y. Liang, J. Li, Y. Wang, M. Zhu, J. Wu, Y. Gao, Y. Jiang, Discontinuous precipitation in a Cu-10Ni-1Si alloy with ultra-high strength, high shock absorption, and good stress relaxation resistance, J. Alloys Compd. 934 (2023) 167676. https://doi.org/10.1016/j.jallcom.2022.167676.

[7] J. Li, G. Huang, X. Mi, L. Peng, H. Xie, Y. Kang, Relationship between the microstructure and properties of a peak aged Cu–Ni–Co–Si alloy, Mater. Sci. Technol. 35 (2019) 606–614. https://doi.org/10.1080/02670836.2019.1576374.

[8] Y. Shan, Y. Zhang, C. Zhang, J. Feng, B. Huang, S. Zhao, K. Song, Study on the microstructure evolution and strengthening mechanism of Cu-Ni-Si alloy after two-stage rolling and aging, Mater. Des. 254 (2025) 113935. https://doi.org/10.1016/j.matdes.2025.113935.

[9] X. Zhang, Q. Liu, X. Liu, X. Zhang, H. Liu, D. Zhao, F. Qi, J. Jie, C. Li, H. Ding, Research on the microstructure and properties of Cu-Ni-Si-Ti alloys with micro-nano hybrid systems, Mater. Des. 259 (2025) 114915. https://doi.org/10.1016/j.matdes.2025.114915.

[10] J.H. Ahn, S.Z. Han, E.-A. Choi, H. Lee, S.H. Lim, J. Lee, K. Kim, N.M. Hwang, H.N. Han, The effect of bimodal structure with nanofibers and normal precipitates on the mechanical and electrical properties of CuNiSi alloy, Mater. Charact. 170 (2020) 110642. https://doi.org/10.1016/j.matchar.2020.110642.

[11] M. Goto, T. Yamamoto, S.Z. Han, T. Utsunomiya, S. Kim, J. Kitamura, J.H. Ahn, S.H. Lim, J. Lee, Simultaneous increase in electrical conductivity and fatigue strength of Cu-Ni-Si alloy by utilizing discontinuous precipitates, Mater. Lett. 288 (2021) 129353. https://doi.org/10.1016/j.matlet.2021.129353.

[12] M. Goto, T. Yamamoto, E.-A. Choi, S. Kim, R. Hirano, S.H. Lim, J.-H. Ahn, J. Lee, S.Z. Han, Physical background of significant increase in mechanical properties and fatigue strength of groove-rolled Cu-Ni-Si alloy with discontinuous precipitates, J. Alloys Compd. 947 (2023) 169569. https://doi.org/10.1016/j.jallcom.2023.169569.

[13] P. Zięba, Recent Developments on Discontinuous Precipitation, Arch. Metall. Mater. Vol. 62, iss. 2A (2017). https://doi.org/10.1515/amm-2017-0138.

[14] X. Meng, G. Xie, W. Xue, Y. Fu, R. Wang, X. Liu, The Precipitation Behavior of a Cu-Ni-Si Alloy with Cr Addition Prepared by Heating-Cooling Combined Mold (HCCM) Continuous Casting, Materials 15 (2022) 4521. https://doi.org/10.3390/ma15134521.

[15] Y. Wu, Y. Li, J. Lu, S. Tan, F. Jiang, J. Sun, Effects of pre-deformation on precipitation behaviors and properties in Cu-Ni-Si-Cr alloy, Mater. Sci. Eng. A 742 (2019) 501–507. https://doi.org/10.1016/j.msea.2018.11.045.
[16] Q.S. Wang, G.L. Xie, X.J. Mi, B.Q. Xiong, X.P. Xiao, The Precipitation and Strengthening Mechanism of Cu-Ni-Si-Co Alloy, Mater. Sci. Forum 749 (2013) 294–298. https://doi.org/10.4028/www.scientific.net/MSF.749.294.
[17] J. Li, G. Huang, X. Mi, L. Peng, H. Xie, Y. Kang, Microstructure evolution and properties of a quaternary Cu–Ni–Co–Si alloy with high strength and conductivity, Mater. Sci. Eng. A 766 (2019) 138390. https://doi.org/10.1016/j.msea.2019.138390.
[18] S. Chenna Krishna, J. Srinath, A.K. Jha, B. Pant, S.C. Sharma, K.M. George, Microstructure and Properties of a High-Strength Cu-Ni-Si-Co-Zr Alloy, J. Mater. Eng. Perform. 22 (2013) 2115–2120. https://doi.org/10.1007/s11665-013-0482-6.
[19] H. Ling, Y. Tang, J. Zhong, X. Meng, M. Zhang, L. Zhang, Thermodynamic, diffusion and precipitation behaviors in Cu–Ni–Si–Co alloys: Modeling and experimental validation, J. Mater. Res. Technol. 35 (2025) 3257–3269. https://doi.org/10.1016/j.jmrt.2025.01.242.
[20] E. Choi, S.Z. Han, J.H. Ahn, S. Semboshi, J. Lee, S.H. Lim, Computational Screening of Efficient Additive Elements to Stabilize the Interface between Cu Matrix and Ni2Si Precipitates in Cu–Ni–Si Alloys, J. Jpn. Inst. Copp. 60 (2021) 293–297. https://doi.org/10.34562/jic.60.1_293.
[21] Y. Zhao, Co-precipitated Ni/Mn shell coated nano Cu-rich core structure: A phase-field study, J. Mater. Res. Technol. 21 (2022) 546–560. https://doi.org/10.1016/j.jmrt.2022.09.032.
[22] S.Z. Han, E.-A. Choi, S.H. Lim, S. Semboshi, Discontinuous Precipitating Behavior for Cu-Ni-Si Alloy with Mn Addition, Mater. Trans. 66 (2025) 23–28. https://doi.org/10.2320/matertrans.MT-D2024006.
[23] W.E. Frazier, T.G. Lach, T.S. Byun, Monte Carlo simulations of Cu/Ni–Si–Mn co-precipitation in duplex stainless steels, Acta Mater. 194 (2020) 1–12. https://doi.org/10.1016/j.actamat.2020.03.053.
[24] S.Z. Han, I.-S. Jeong, B. Ryu, S.J. Lee, J.H. Ahn, E.-A. Choi, Enhanced strength of Cu-Ni-Si alloy via heterogeneous nucleation at grain boundaries during homogenization, Mater. Charact. 215 (2024) 114198. https://doi.org/10.1016/j.matchar.2024.114198.
[25] H. Liu, S. Wen, Y. Liu, C. Du, Q. Min, Z. Chen, Y. Du, Diffusion kinetics for fcc quinary Cu–Co–Mn–Ni–Si system and its application to precipitation simulations, J. Mater. Res. Technol. 24 (2023) 675–688. https://doi.org/10.1016/j.jmrt.2023.03.033.
[26] I. Shuro, H.H. Kuo, T. Sasaki, K. Hono, Y. Todaka, M. Umemoto, G-phase precipitation in austenitic stainless steel deformed by high pressure torsion, Mater. Sci. Eng. A 552 (2012) 194–198. https://doi.org/10.1016/j.msea.2012.05.030.
[27] M. Murayama, K. Hono, Y. Katayama, Microstructural evolution in a 17-4 PH stainless steel after aging at 400 °C, Metall. Mater. Trans. A 30 (1999) 345–353. https://doi.org/10.1007/s11661-999-0323-2.
[28] J. Wang, H. Zou, C. Li, S. Qiu, B. Shen, The effect of microstructural evolution on hardening behavior of type 17-4PH stainless steel in long-term aging at 350 °C, Mater. Charact. 57 (2006) 274–280. https://doi.org/10.1016/j.matchar.2006.02.004.
[29] D. Kleiven, J. Akola, Precipitate formation in aluminium alloys: Multi-scale modelling approach, Acta Mater. 195 (2020) 123–131. https://doi.org/10.1016/j.actamat.2020.05.050.
[30] H.M. Nam, H. Lee, H. Kim, J.G. Lee, J. Kim, Precipitation strengthening of Cu–Ni–Si-based alloys: Experimental and computational insights, J. Mater. Res. Technol. 27 (2023) 5372–5379. https://doi.org/10.1016/j.jmrt.2023.11.019.

[31] H. Peng, C. Zhou, W. Xie, H. Chen, Interfacial characteristics and their influence on stress relaxation in Cu-Ni-Co-Si alloys, J. Alloys Compd. 1038 (2025) 182688. https://doi.org/10.1016/j.jallcom.2025.182688.
[32] S. Hocker, H. Lipp, E. Eisfeld, S. Schmauder, J. Roth, Precipitation strengthening in Cu–Ni–Si alloys modeled with ab initio based interatomic potentials, J. Chem. Phys. 149 (2018) 024701. https://doi.org/10.1063/1.5029887.
[33] Y. Mishin, Machine-learning interatomic potentials for materials science, Acta Mater. 214 (2021) 116980. https://doi.org/10.1016/j.actamat.2021.116980.
[34] T. Mueller, A. Hernandez, C. Wang, Machine learning for interatomic potential models, J. Chem. Phys. 152 (2020) 050902. https://doi.org/10.1063/1.5126336.
[35] A.P. Bartók, J. Kermode, N. Bernstein, G. Csányi, Machine Learning a General-Purpose Interatomic Potential for Silicon, Phys. Rev. X 8 (2018) 041048. https://doi.org/10.1103/PhysRevX.8.041048.
[36] B. Mortazavi, X. Zhuang, T. Rabczuk, A.V. Shapeev, Atomistic modeling of the mechanical properties: the rise of machine learning interatomic potentials, Mater. Horiz. 10 (2023) 1956–1968. https://doi.org/10.1039/D3MH00125C.
[37] G. Wang, C. Wang, X. Zhang, Z. Li, J. Zhou, Z. Sun, Machine learning interatomic potential: Bridge the gap between small-scale models and realistic device-scale simulations, iScience 27 (2024) 109673. https://doi.org/10.1016/j.isci.2024.109673.
[38] Y.-W. Zhang, V. Sorkin, Z.H. Aitken, A. Politano, J. Behler, A. P Thompson, T.W. Ko, S.P. Ong, O. Chalykh, D. Korogod, E. Podryabinkin, A. Shapeev, J. Li, Y. Mishin, Z. Pei, X. Liu, J. Kim, Y. Park, S. Hwang, S. Han, K. Sheriff, Y. Cao, R. Freitas, Roadmap for the development of machine learning-based interatomic potentials, Model. Simul. Mater. Sci. Eng. 33 (2025) 023301. https://doi.org/10.1088/1361-651X/ad9d63.
[39] N. Leimeroth, L.C. Erhard, K. Albe, J. Rohrer, Machine-learning interatomic potentials from a users perspective: a comparison of accuracy, speed and data efficiency, Model. Simul. Mater. Sci. Eng. 33 (2025) 065012. https://doi.org/10.1088/1361-651X/adf56d.
[40] V.L. Deringer, M.A. Caro, G. Csányi, Machine Learning Interatomic Potentials as Emerging Tools for Materials Science, Adv. Mater. 31 (2019) 1902765. https://doi.org/10.1002/adma.201902765.
[41] W. Yu, C. Ji, X. Wan, Z. Zhang, J. Robertson, S. Liu, Y. Guo, Machine-learning-based interatomic potentials for advanced manufacturing, Int. J. Mech. Syst. Dyn. 1 (2021) 159–172. https://doi.org/10.1002/msd2.12021.
[42] N. Artrith, A. Urban, G. Ceder, Efficient and accurate machine-learning interpolation of atomic energies in compositions with many species, Phys. Rev. B 96 (2017) 014112. https://doi.org/10.1103/PhysRevB.96.014112.
[43] P. Hohenberg, W. Kohn, Inhomogeneous Electron Gas, Phys. Rev. 136 (1964) B864–B871. https://doi.org/10.1103/PhysRev.136.B864.
[44] W. Kohn, L.J. Sham, Self-Consistent Equations Including Exchange and Correlation Effects, Phys. Rev. 140 (1965) A1133–A1138. https://doi.org/10.1103/PhysRev.140.A1133.
[45] J.P. Perdew, K. Burke, M. Ernzerhof, Generalized Gradient Approximation Made Simple, Phys. Rev. Lett. 77 (1996) 3865–3868. https://doi.org/10.1103/PhysRevLett.77.3865.
[46] P.E. Blöchl, Projector augmented-wave method, Phys. Rev. B 50 (1994) 17953–17979. https://doi.org/10.1103/PhysRevB.50.17953.
[47] G. Kresse, J. Furthmüller, Efficient iterative schemes for ab initio total-energy calculations using a plane-wave basis set, Phys. Rev. B 54 (1996) 11169–11186. https://doi.org/10.1103/PhysRevB.54.11169.
[48] C. Chen, S.P. Ong, A universal graph deep learning interatomic potential for the periodic table, Nat. Comput. Sci. 2 (2022) 718–728. https://doi.org/10.1038/s43588-022-00349-3.

[49] Y. Park, J. Kim, S. Hwang, S. Han, Scalable Parallel Algorithm for Graph Neural Network Interatomic Potentials in Molecular Dynamics Simulations, J. Chem. Theory Comput. 20 (2024) 4857–4868. https://doi.org/10.1021/acs.jctc.4c00190.
[50] J. Kim, J. You, Y. Park, Y. Lim, Y. Kang, J. Kim, H. Jeon, S. Ju, D. Hong, S.Y. Lee, S. Choi, Y. Kim, J.W. Lee, S. Han, Optimizing cross-domain transfer for universal machine learning interatomic potentials, Nat. Commun. (2026). https://doi.org/10.1038/s41467-026-70195-8.
[51] B. Deng, P. Zhong, K. Jun, J. Riebesell, K. Han, C.J. Bartel, G. Ceder, CHGNet as a pretrained universal neural network potential for charge-informed atomistic modelling, Nat. Mach. Intell. 5 (2023) 1031–1041. https://doi.org/10.1038/s42256-023-00716-3.
[52] J. Kim, J. Kim, J. Kim, J. Lee, Y. Park, Y. Kang, S. Han, Data-Efficient Multifidelity Training for High-Fidelity Machine Learning Interatomic Potentials, J. Am. Chem. Soc. 147 (2025) 1042–1054. https://doi.org/10.1021/jacs.4c14455.
[53] D. Xie, S.B. Biggers, Calculation of transient strain energy release rates under impact loading based on the virtual crack closure technique, Int. J. Impact Eng. 34 (2007) 1047–1060. https://doi.org/10.1016/j.ijimpeng.2006.02.007.
[54] S.Y. Hu, M.I. Baskes, M. Stan, L.Q. Chen, Atomistic calculations of interfacial energies, nucleus shape and size of $\theta'$ precipitates in Al–Cu alloys, Acta Mater. 54 (2006) 4699–4707. https://doi.org/10.1016/j.actamat.2006.06.010.
[55] T. Yang, X. Chen, W. Li, X. Han, P. Liu, First-principles calculations to investigate the interfacial energy and electronic properties of Mg/AlN interface, J. Phys. Chem. Solids 167 (2022) 110705. https://doi.org/10.1016/j.jpcs.2022.110705.
[56] N.M. Stuart, K. Sohlberg, A method of calculating surface energies for asymmetric slab models, Phys. Chem. Chem. Phys. 25 (2023) 13351–13358. https://doi.org/10.1039/D2CP04460A.
[57] V. Fiorentini, M. Methfessel, Extracting convergent surface energies from slab calculations, J. Phys. Condens. Matter 8 (1996) 6525. https://doi.org/10.1088/0953-8984/8/36/005.
[58] D.A. Porter, K.E. Easterling, K.E. Easterling, Phase Transformations in Metals and Alloys (Revised Reprint), 3rd ed., CRC Press, Boca Raton, 2009. https://doi.org/10.1201/9781439883570.
[59] J.W. Christian, The Theory of Transformations in Metals and Alloys, Pergamon, 2007. https://doi.org/10.1016/B978-0-08-044019-4.X5000-4.
[60] X. Yan, A. Grytsiv, P. Rogl, V. Pomjakushin, X. Xue, On the crystal structure of the Mn–Ni–Si G-phase, J. Alloys Compd. 469 (2009) 152–155. https://doi.org/10.1016/j.jallcom.2008.01.142.
[61] C. Watanabe, R. Monzen, Coarsening of δ-Ni2Si precipitates in a Cu–Ni–Si alloy, J. Mater. Sci. 46 (2011) 4327–4335. https://doi.org/10.1007/s10853-011-5261-x.

# Supplementary Materials

## Precipitate phase selection and grain boundary morphology in Cu-Ni-Si-Mn alloys: A machine-learning interatomic potential study

Aadil Fayaz Wani,[1] Il-Seok Jeong,[2] Haekwan Jeon,[3] Jaesun Kim,[3] SuDong Park,[1] Eun-Ae Choi,[2,4*] Seung Zeon Han,[2,4] Seungwu Han,[3,5] and Byungki Ryu[1,6*]

7- Energy Conversion Research Center, Electrical Materials Research Division, Korea Electrotechnology Research Institute (KERI), Changwon-si 51543, Republic of Korea
8- Extreme Materials Research Institute, Korea Institute of Materials Science (KIMS), Changwon-si 51508, Republic of Korea
9- Department of Materials Science and Engineering, Seoul National University, Seoul 08826, Republic of Korea
10- Advanced Materials Engineering, KIMS School, University of Science and Technology (UST), Changwon-si 51508, Republic of Korea
11- AI Center, Korea Institute for Advanced Study, Seoul 02455, Republic of Korea
12- Electrical-Functionality Materials Engineering, KERI School, University of Science and Technology (UST), Changwon-si 51543, Republic of Korea

*Correspondence: byungkiryu@keri.re.kr (BR) eunae.choi@kims.re.kr (EA)

**Table S1.** Optimized structural parameters of cubic Cu and orthorhombic δ-Ni2Si obtained using DFT and different MLIP models for interface construction

| Model | axis | Cu (Å) | Relaxed-$Ni_2Si$ (Å) | $Ni_2Si$-on-Cu (Å) | Error Cu (%) | Error $Ni_2Si$ (%) |
|---|---|---|---|---|---|---|
| **DFT** | a | 15.405 | 14.123 | 14.241 | | |
| | b | 5.135 | 4.993 | 5.135 | | |
| | c | 3.631 | 3.724 | 3.631 | | |
| **M3GNet** | a | 15.326 | 14.408 | 14.408 | 0.513 | -2.018 |
| | b | 5.109 | 4.942 | 5.109 | 0.506 | 1.021 |
| | c | 3.612 | 3.695 | 3.612 | 0.523 | 0.779 |
| **7net-0** | a | 15.325 | 14.183 | 14.238 | 0.519 | -0.425 |
| | b | 5.108 | 4.967 | 5.108 | 0.526 | 0.521 |
| | c | 3.631 | 3.729 | 3.613 | 0.496 | -0.134 |
| **7net-Ompa-mpa** | a | 15.732 | 14.439 | 14.208 | 0.214 | -0.326 |
| | b | 5.124 | 4.994 | 5.124 | 0.214 | -0.020 |
| | c | 3.623 | 3.725 | 3.623 | 0.220 | -0.027 |
| **7net-Ompa-omat** | a | 15.243 | 14.145 | 14.209 | 1.052 | -0.156 |
| | b | 5.124 | 4.985 | 5.114 | 0.409 | 0.160 |
| | c | 3.623 | 3.719 | 3.616 | 0.413 | 0.134 |
| **7net-Omni-mpa** | a | 15.243 | 14.194 | 14.210 | 0.221 | -0.503 |
| | b | 5.114 | 4.996 | 5.124 | 0.214 | -0.060 |
| | c | 3.616 | 3.710 | 3.623 | 0.220 | 0.376 |
| **7net-Omni-omat** | a | 15.344 | 14.201 | 14.224 | 0.396 | -0.545 |
| | b | 5.114 | 4.989 | 5.114 | 0.09 | 0.080 |
| | c | 3.616 | 3.709 | 3.616 | 0.413 | 0.403 |

**Table S2.** Optimized structural parameters of orthorhombic Cu and cubic G-phase obtained using DFT and different MLIP models for interface construction

| Model | axis | Cu (Å) | Relaxed-G (Å) | G-on-Cu (Å) | Error Cu (%) | Error G (%) |
|---|---|---|---|---|---|---|
| DFT | a | 7.701 | 7.826 | 7.701 | | |
| | b | 7.701 | 7.826 | 7.701 | | |
| | c | 10.893 | 11.07 | 11.085 | | |
| M3GNet | a | 7.663 | 7.830 | 7.664 | 0.498 | −0.045 |
| | b | 7.663 | 7.830 | 7.664 | 0.498 | −0.045 |
| | c | 10.837 | 11.073 | 11.390 | 0.514 | -0.027 |
| 7net-0 | a | 7.664 | 7.834 | 7.664 | 0.485 | -0.096 |
| | b | 7.664 | 7.834 | 7.664 | 0.485 | -0.096 |
| | c | 10.839 | 11.078 | 11.359 | 0.496 | -0.072 |
| 7net-Ompa-mpa | a | 7.685 | 7.841 | 7.685 | 0.212 | -0.185 |
| | b | 7.685 | 7.841 | 7.685 | 0.212 | -0.185 |
| | c | 10.869 | 11.089 | 11.317 | 0.220 | -0.172 |
| 7net-Ompa-omat | a | 7.676 | 7.833 | 7.672 | 0.381 | -0.083 |
| | b | 7.676 | 7.833 | 7.672 | 0.381 | -0.083 |
| | c | 10.849 | 11.077 | 11.360 | 0.404 | -0.063 |
| 7net-Omni-mpa | a | 7.683 | 7.844 | 7.683 | 0.238 | -0.024 |
| | b | 7.683 | 7.844 | 7.683 | 0.238 | -0.024 |
| | c | 10.86 | 11.093 | 11.307 | 0.248 | -0.208 |
| 7net-Omni-omat | a | 7.672 | 9.836 | 7.672 | 0.381 | -0.122 |
| | b | 7.672 | 7.836 | 7.672 | 0.381 | -0.122 |
| | c | 10.849 | 11.082 | 11.298 | 0.404 | -0.108 |

**Table S3.** Calculated values of strain energy ($E_{\mathrm{str}}$), surface energy ($E_{\mathrm{SF}}$), and interface energy ($E_{\mathrm{IF}}$) of $\delta$-$Ni_2Si$ and G-phase calculated using DFT and various machine learning interatomic potential (MLIP) models.

| ***Model*** | $\boldsymbol{E_{\mathrm{str}}}$ $(\mathbf{eV \cdot nm^{-3}})$ | | $\boldsymbol{E_{SF}}$ $(\mathbf{J \cdot m^{-2}})$ | | $\boldsymbol{E_{IF}}$ $(\mathbf{J \cdot m^{-2}})$ | |
|---|---|---|---|---|---|---|
| | **$Ni_2Si$** | **G-phase** | **$Ni_2Si$** | **G-phase** | **$Ni_2Si$** | **G-phase** |
| DFT | 0.59 | 0.13 | 2.01 | 2.26 | 0.48 | 0.98 |
| M3GNet | 0.12 | 0.63 | 1.07 | 1.15 | 0.32 | 0.73 |
| 7net-0 | 0.43 | 0.80 | 1.73 | 1.81 | 0.62 | 1.21 |
| 7net-Ompa-mpa | 0.50 | 0.73 | 2.07 | 2.15 | 0.45 | 0.98 |
| 7net-Ompa-omat | 0.52 | 0.78 | 2.11 | 2.18 | 0.45 | 0.96 |
| 7net-Omni-mpa | 0.42 | 0.78 | 2.03 | 2.10 | 0.53 | 0.88 |
| 7net-Omni-omat | 0.48 | 1.17 | 2.07 | 2.12 | 0.50 | 0.85 |

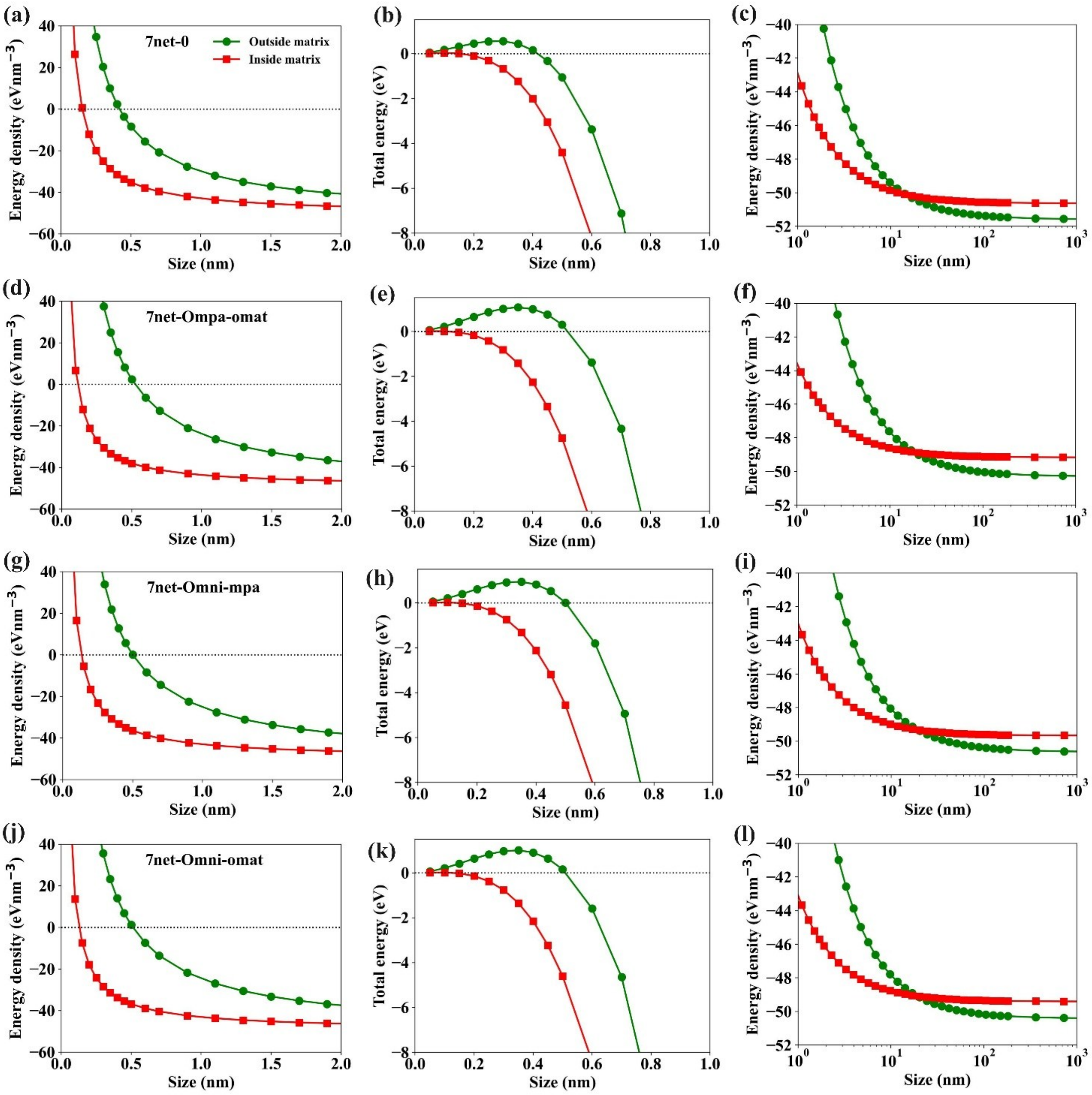

**Figure S1.** Comparison of energy density (first and third column) and total energy (second column) predicted using 7net-0, 7net-Ompa-omat, 7net-Omni-mpa, and 7net-Omni-omat MLIP models at small and large particle sizes of $Ni_2Si$ precipitates inside and outside the Cu matrix.

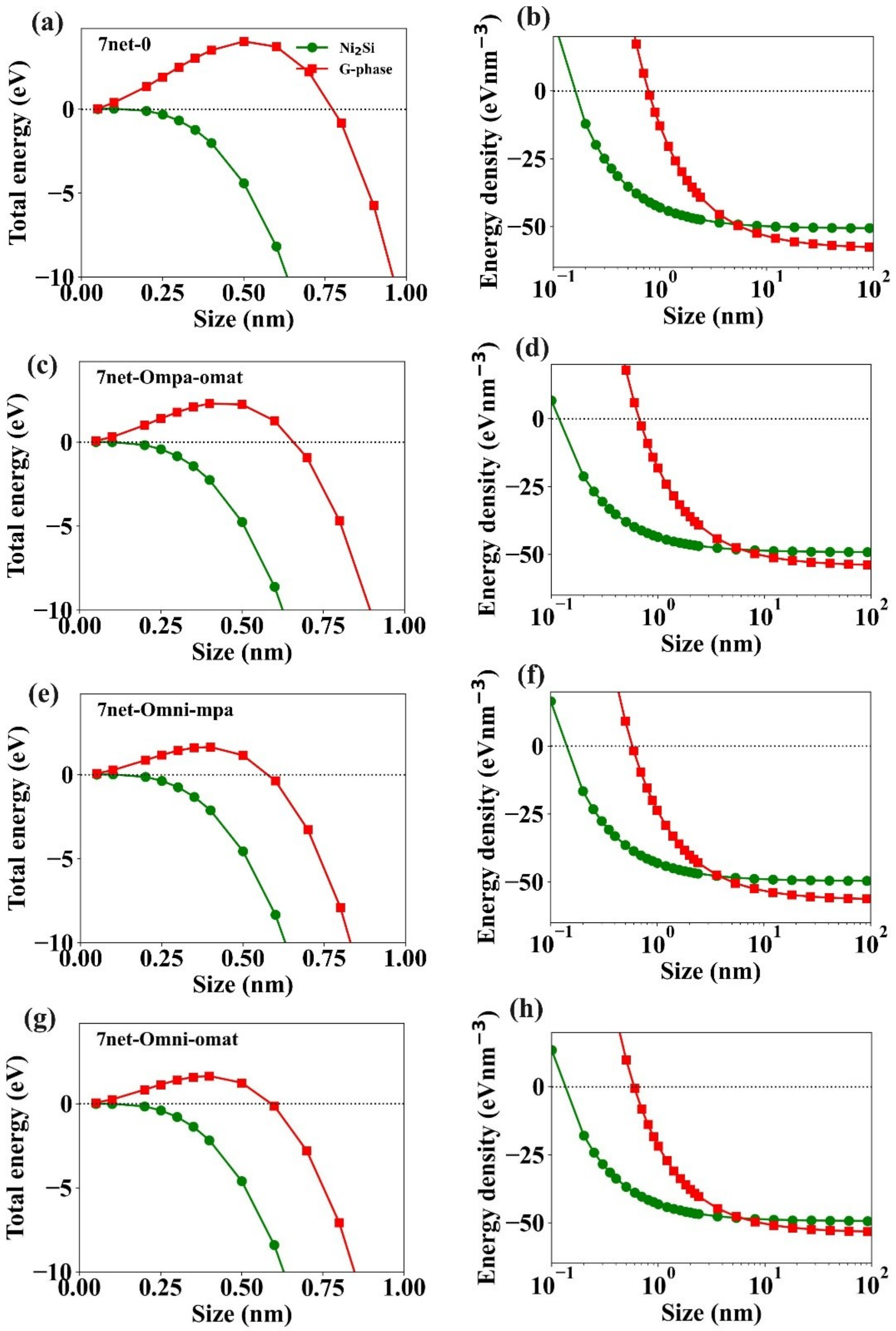

**Figure S2.** Size-dependent comparison of total energy (first column) and energy density (second column) of $Ni_2Si$ and G-phase computed using 7net-0, 7net-Ompa-omat, 7net-Omni-mpa, and 7net-Omni-omat MLIP models.